%% file: 3DTBL.tex
\documentclass{jfm}

\usepackage{hyperref,xcolor}
\usepackage{natbib}
\usepackage{amsbsy}
\usepackage{psfrag}
\usepackage{amsmath}
\usepackage{soul}
\usepackage{subfig}
\usepackage{etoolbox}

\input{macros.tex}

\hypersetup{
  colorlinks=true,
  linkcolor=blue,
  citecolor=blue
}

\usepackage[labelsep=period]{caption}
\renewcommand{\tablename}{\sc Table}
\renewcommand{\figurename}{\sc Figure}

\makeatletter

\let\oldciteauthor\citeauthor

\def\citeauthor#1{{\NoHyper\oldciteauthor{#1}}}

\patchcmd{\NAT@citex}
  {\@citea\NAT@hyper@{%
     \NAT@nmfmt{\NAT@nm}%
     \hyper@natlinkbreak{\NAT@aysep\NAT@spacechar}{\@citeb\@extra@b@citeb}%
     \NAT@date}}
  {\@citea\NAT@nmfmt{\NAT@nm}%
   \NAT@aysep\NAT@spacechar\NAT@hyper@{\NAT@date}}{}{}

\patchcmd{\NAT@citex}
  {\@citea\NAT@hyper@{%
     \NAT@nmfmt{\NAT@nm}%
     \hyper@natlinkbreak{\NAT@spacechar\NAT@@open\if*#1*\else#1\NAT@spacechar\fi}%
       {\@citeb\@extra@b@citeb}%
     \NAT@date}}
  {\@citea\NAT@nmfmt{\NAT@nm}%
   \NAT@spacechar\NAT@@open\if*#1*\else#1\NAT@spacechar\fi\NAT@hyper@{\NAT@date}}
  {}{}

\makeatother

\title{Non-equilibrium three-dimensional boundary layers at moderate Reynolds numbers}
\shorttitle{Non-equilibrium 3-D boundary layers}
\shortauthor{A. Lozano-Dur\'an, M. Giometto, G.~I. Park, \& P. Moin}

\author{
Adri\'an Lozano-Dur\'an$^1$\corresp{\email{adrianld@stanford.edu}},  
Marco Giometto$^{2}$, 
George I. Park$^{3}$, \& 
Parviz Moin$^{1}$
}
\affiliation{
$^1$Center for Turbulence Research, Stanford University, California 94305, USA \\[\affilskip]
$^{2}$Department of Civil Engineering and Engineering Mechanics, Columbia University, New York, 10027, USA\\[\affilskip]
$^{3}$Department of Mechanical Engineering and Applied Mechanics, University
of Pennsylvania, Philadelphia, Pennsylvania 19104, USA 
}

\begin{document}

\maketitle

\begin{abstract}
Non-equilibrium wall turbulence with mean-flow three-dimensionality is
ubiquitous in geophysical and engineering flows.  Under these
conditions, turbulence may experience a counter-intuitive depletion of
the turbulent stresses, which has important implications for modelling
and control. Yet, current turbulence theories have been established
mainly for statistically two-dimensional equilibrium flows and are
unable to predict the reduction in the Reynolds stress magnitude. In
the present work, we propose a multiscale model which explains the
response of non-equilibrium wall-bounded turbulence under the
imposition of three-dimensional strain. The analysis is performed via
direct numerical simulation of transient three-dimensional turbulent
channels subjected to a sudden lateral pressure gradient at friction
Reynolds numbers up to 1,000. We show that the flow regimes and
scaling properties of the Reynolds stress are consistent with a model
comprising momentum-carrying eddies with sizes and time scales
proportional to their distance to the wall. We further demonstrate
that the reduction in Reynolds stress follows a spatially and
temporally self-similar evolution caused by the relative horizontal
displacement between the core of the momentum-carrying eddies and the
flow layer underneath. Inspection of the flow energetics reveals that
this mechanism is associated with lower levels of pressure-strain
correlation which ultimately inhibits the generation of Reynolds
stress. Finally, we assess the ability of the state-of-the-art
wall-modelled large-eddy simulation to predict non-equilibrium,
three-dimensional flows.
\end{abstract}


\section{Introduction}


Our current understanding of wall turbulence is largely rooted in
studies of equilibrium boundary layers with two-dimensional (2-D) mean
velocity profiles (i.e., contained in a plane). However,
non-equilibrium turbulence with mean-flow three-dimensionality is the
rule rather than the exception in most geophysical and engineering
flows.  Prominent examples of the former are Ekman layers and spirals,
flow in complex terrain, tornadoes, and river bends, while industrial
flows include flow over swept-wing aircrafts and hulls of marine
vehicles, around buildings and obstacles, within turbomachines,
etc. Despite the ubiquity of such flows, fundamental questions remain
unanswered regarding the structural changes of wall turbulence under
three-dimensional (3-D) non-equilibrium conditions, challenging our
intellectual ability to comprehend and predict wall turbulence in
broader scenarios. In the present work, we study the transition of
statistically stationary 2-D turbulence to non-stationary 3-D states
induced by the sudden application of a spanwise pressure gradient.
Our emphasis is on the multiscale structure of wall-bounded turbulence
at moderately high Reynolds numbers.

The vast majority of the fundamental studies on wall turbulence has
focused on a narrow subset of equilibrium 2-D wall-bounded flows
(2DTBL) such as turbulent channels \citep{Kim1987,Lee2015}, pipes
\citep{Wu2015,Pirozzoli2018}, and flat plates boundary layers
\citep{Spalart1988,Sillero2013,Sillero2014,Wu2017}. These studies have
unravelled constitutive characteristics of the near-wall turbulence,
including its self-sustaining nature \citep{Jimenez1991,Jimenez1999,
  Panton2001, Flores2010, Hwang2011, Hwang2015, Farrell2016,
  Farrell2017}, the coherent structure and geometry of the flow
\citep{DelAlamo2006,Kawahara2012,Lozano2012,Dong2017,McKeon2017}, the
life cycle of the momentum-carrying eddies \citep{Lozano2014b,
  Hwang2010, Cossu2017}, and the wall-attached structure of the flow
in the logarithmic layer \citep{Marusic2013, Hwang2016, Chandran2017,
  Marusic2019, Cheng2019}, among others. Unfortunately, theories built
upon equilibrium wall-turbulence have had limited impact on our
ability to predict 3-D boundary layers (3DTBL) and to grasp the
physics underlying the extensive collection of numerical and
experimental observations. This is principally due to the violation of
the temporal/spatial homogeneity of the flow and the unidirectionality
of the mean shear, which are foundational assumptions of 2DTBL absent
in 3DTBL. Consequently, the knowledge established largely for
equilibrium 2DTBL, such as the law-of-the-wall \citep{Prandtl1925,
  Millikan1938, Coles1969}, the scaling laws for the velocity and
energy spectra \citep{Perry1975, Perry1977, Zagarola1998,
  Morrison2004, DelAlamo2004,Marusic2013, Vallikivi2015, Hoyas2006,
  Klewicki2007, Chandran2017}, structural models of the flow
\citep{Townsend1976, Adrian2000, Meneveau2013, Agostini2017,
  Lozano2019, Jimenez2018, Marusic2019}, and reduced-order models
\citep{Rowley2017,Durbin2018,Bose2018}, cannot be generalised
trivially to non-canonical 3DTBL.

Often, 3DTBL are classified according to their state as either in
equilibrium or in non-equilibrium.  \cite{Townsend1961} was the first
to coin the term `equilibrium layer' to define a portion of the
boundary layer in which the rates of production and dissipation of
turbulent kinetic energy are equal.  \cite{DeGraaff2000} suggested a
more restrictive definition where the total shear stress is balanced
by the shear stress at the wall. A comprehensive theory of equilibrium
and self-similar flow motions in the outer region of turbulent
boundary layers can be also found in the works by \cite{Castillo2001}
and \cite{Maciel2006,Maciel2018}.  Here, we refer to equilibrium flow
simply as that in statistically stationary state. Despite equilibrium
3DTBL, such as the Ekman layer, are of paramount importance (see
e.g. \cite{Spalart1989, Coleman1990,Littell1994, Wu1997,
  Coleman2000}), the subject of the present work is the
non-equilibrium response of 3DTBL, which is one of the most
challenging cases for the current turbulence theories.  In addition to
their equilibrium state (or lack thereof), 3DTBL are also classified
according to the mechanisms by which the three-dimensionality is
incorporated into the flow. In this respect, 3DTBL can be labelled as
`viscous-induced' when the three-dimensionality is a direct
consequence of the viscous effects propagating from the solid
boundaries (e.g., moving walls, accelerating frames of reference,...),
or as `inviscid-induced' when the 3-D flow is the result of
space-varying body forces or pressure gradients (such as those
triggered by the presence of complex geometries or by baroclinic
effects in atmospheric flows). These two mechanisms are usually
referred to as pressure-driven and shear-driven in the literature,
although such a nomenclature may lead to confusion in some situations.
Here we are concerned with the first kind, i.e. `viscous-induced'
3DTBL, which are relevant for turbomachinery applications and
large-scale wind farms, just to mention two examples, albeit it is
worth noting that in many real life scenarios three-dimensionality is
induced by a combination of the two mechanisms.

From the early works by \cite{Bradshaw1969} and \cite{Berg1972}, it
was readily noted that 3DTBL exhibit a response contrary to the common
expectations from their 2-D counterparts. Such counter-intuitive
effects manifest themselves in the reduction of the tangential
Reynolds stress and the misalignment of the Reynolds stress and mean
shear vectors. These observations have been reported for both
equilibrium and non-equilibrium 3DTBL, albeit the effects are
exacerbated in the latter. The pioneering studies on 3DTBL were
laboratory experiments. \cite{Bradshaw1969} presented the first set of
Reynolds stress measurements in an yawed flat plate as a surrogate of
an `infinite' swept wing. They observed a lag between the Reynolds
stress angle and the mean velocity gradient angle despite the mild
three-dimensionality of the flow. Subsequent experiments by
\cite{Johnston1970}, \cite{Berg1975} and \cite{Bradshaw1985} confirmed
the aforementioned behaviour in similar set-ups.  In a succeeding
series of studies, \cite{Berg1972}, \cite{Elsenaar1974} and
\cite{Berg1975} further showed that the intensity of the Reynolds
stress for a given amount of turbulent kinetic energy
(a.k.a. Townsend's structure parameter) dropped below the commonly
reported value in 2-D flows, establishing the second main
counter-intuitive effect of 3DTBL.

Over the past decades, a variety of additional experimental studies on
3DTBL have been performed, each characterised by the different
mechanism utilised to induce three-dimensionality in the flow. Among
them, we can highlight 3DTBL over wedges
\citep{Anderson1987,Anderson1989,Compton1997}, rotating cylinders
\citep{Furuya1966, Bissonnette1974,
  Lohmann1976,Driver1987,Driver1989,Driver1991}, rotating disks
\citep{Littell1994}, flow within the bend of ducts
\citep{Schwarz1993,Schwarz1994,Flack1993,Flack1994}, swept steps and
bumps \citep{Flack1993,Webster1996}, and wing-body junctions
\citep{Olcmen1992,Olcmen1995}.  More recently,
\cite{Kiesow2002,Kiesow2003} used particle-image velocimetry (PIV) to
acquire detailed information of the flow structure at varying degrees
of cross-flow generated by moving belts.  The large body of literature
on experimental 3DTBL until the 1990s is summarised in the reviews by
\cite{Fernholz1981}, \cite{Berg1988}, \cite{Eaton1995} and
\cite{Johnston1996}.

The advent of direct numerical simulation (DNS) and large-eddy
simulation (LES) led to an increase in the number of numerical
investigations of 3DTBL.  Computational studies carried out to date
include channel flows subject to transverse pressure gradients
\citep{Moin1990,Sendstad1992,Coleman1996a,He2018}, flat plates with
time-dependent free-stream velocity \citep{Spalart1989}, rotating
disks \citep{Littell1994,Wu2000}, Couette flows with spanwise pressure
gradient \citep{Holstad2010}, and concentric annulus with rotating
inner wall \citep{Jung2006}, among others. \cite{Coleman1996a,
  Coleman1996b, Coleman2000} computed DNS of initially 2-D
fully-developed turbulence subjected to mean strains, emulating the
effect of rapid spatially-varying changes of the pressure gradients in
ducts or diffusers.  \cite{Wu1997,Wu1998} performed LES of the swept
bump proposed experimentally by \cite{Webster1996}, while other
numerical investigations have introduced three-dimensionality in flow
by the impulsive motion of walls in the spanwise direction
\citep{Howard1997, Le1999a, Le1999b}, by spanwise oscillating walls
\citep{Jung1992}, and by a sustained lateral displacement of a finite
section of the wall \citep{Kannepalli2000}.



The current consensus among the experimental and numerical studies
above is that three-dimensionality of the mean flow is typically
accompanied by a decrease of the tangential Reynolds stress, the
reduction of drag, and the misalignment of the mean Reynolds stress
vector and mean shear vector.  Given that equilibrium 2-D turbulence
is commonly enhanced by the addition of mean shear, the previous
results are non-trivial to interpret. Accordingly, there have been
multiple attempts to reconcile the non-intuitive flow response with
the traditional structural organisation of near-wall turbulence
\citep{Jimenez1991, Jimenez1999, Schoppa2002}. Most structural studies
of 3DTBL depart from the premise that 2DTBL are structurally `optimal'
for the generation of Reynolds stress, and that 3DTBL are essentially
a distorted, less efficient version of the former. \cite{Lohmann1976}
postulated one of the first structural pictures of the flow by
suggesting that transverse shear was responsible for the break up of
quasi-streamwise vortices into smaller structures. \cite{Bradshaw1985}
further hypothesised that eddies were tilted away from their preferred
alignment by the spanwise strain, which impeded the production of
Reynolds stress. \cite{Eaton1991} stated that low-speed streaks are
inhibited by the mean cross-flow, which reduces the number of
ejections (and hence of Reynolds stress) generated via streak
instability and breakdown.  \cite{Kannepalli2000} also observed
significant disruption of the near-wall streaks at both the leading
and trailing edge of the moving wall section as the flow adjusts to
the new wall boundary conditions. Later PIV measurements by
\cite{Kiesow2002} confirmed a significant alteration of the near-wall
flow physics, with significant disruption of the streak length
compared to 2DTBL. On the other hand, the works by
\cite{Anderson1989}, \cite{Sendstad1992}, \cite{Littell1994},
\cite{Eaton1995}, and \cite{Chiang1996}, have centred the attention on
the strong asymmetry between vortices of different sign rather than on
streaks as the main cause for stress reduction. They argued that the
intrinsic structure of 3DTBL favour either a sweep or a ejection,
which reduces the efficiency of the boundary layer to produce Reynolds
stress. The LES by \cite{Wu1997} supported the structural model
proposed by \cite{Littell1994}. However, \cite{Jung2006} rendered the
latter scenario invalid in a concentric annulus by analysing the
distinctive flow features using conditional analysis.

Finally, it is worth mentioning that the peculiarities of 3DTBL are
expected to undermine the performance of modelling techniques built on
and validated for 2DTBL. Especially concerning is the development and
testing of wall models for LES, motivated by the need to bypass the
inner wall region in order to reduce computational costs
\citep{Chapman1979,Choi2012}.  Early wall models relying on
equilibrium assumptions have yielded fair predictions in simple flows,
but are known to be suboptimal in more complex configurations
\citep{Larsson2016}.  This has motivated recent efforts to develop new
wall models accounting for non-equilibrium effects
\citep{Balaras1996,Wang2002,Yang2015,Park2014}, free of tunable
parameters \citep{Bose2014,Lozano2017,Bae2018b}, and capable of
delivering robust predictions for non-canonical flow settings
\citep[see for instance the recent review by][]{Bose2018}.  Note that,
in general, wall models are not effective at transferring information
of the flow structure from the inner to the outer layer
\citep{Piomelli2002}.  Hence, the current flow set-up characterised by
a spanwise boundary layer growing from the wall is a challenging
testbed for wall-model LES (WMLES).

The primary foci of this work are the investigation of the scaling
properties of 3DTBL, absent in previous numerical studies at low
Reynolds numbers, and the elucidation of the structural mechanisms
responsible for Reynolds stress deficit during the initial transient.
The insight gained in used to envision a multiscale structural model
consistent with the scalings and structural changes observed. We also
inspect the implications of three-dimensionality and non-equilibrium
state for WMLES. A preliminary version of this work can be found in
\citet{Giometto2017}. The paper is organised as follows. The numerical
set-up and database are presented in \S\ref{sec:numerics}. The
analysis of the scaling and flow structure of the flow is discussed in
\S\ref{sec:analysis}. In \S\ref{sec:wmles}, we focus on the comparison
of selected quantities for DNS and wall-modelled LES. Finally,
conclusions are offered in \S\ref{sec:conclusions}.

\section{Problem set-up and numerical database}\label{sec:numerics}

\begin{figure}
\vspace{0.5cm}
\begin{center}
\includegraphics[width=0.9\textwidth]{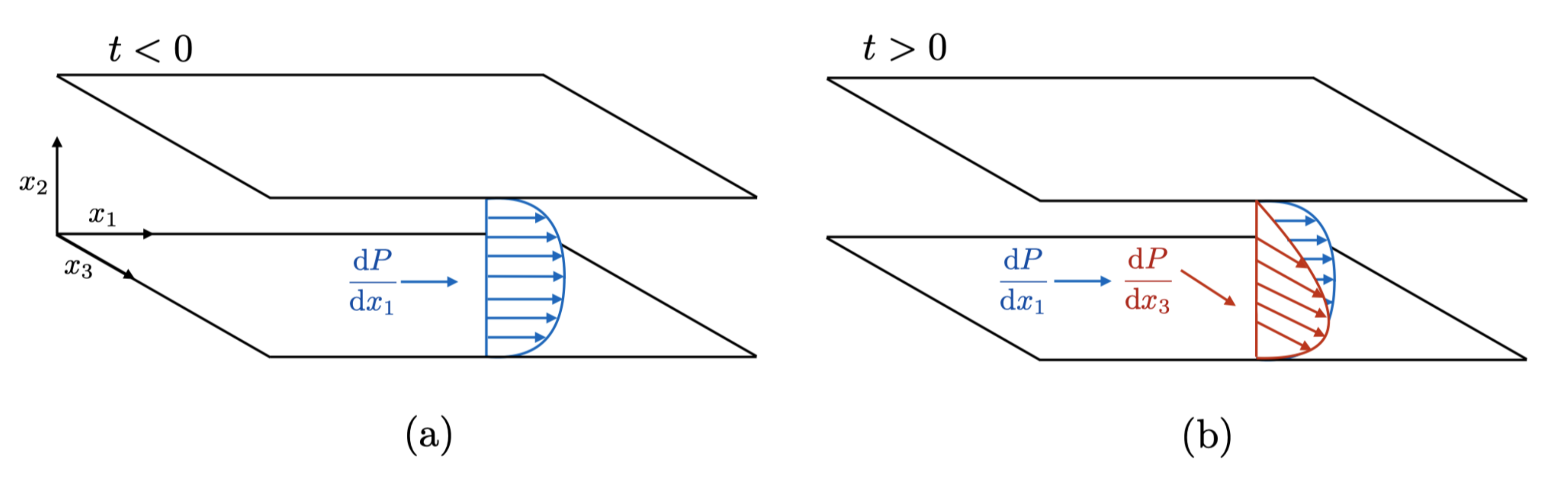}
\caption{ Schematic of the numerical set-up of a 2-D fully-developed
  turbulent channel flow subjected to a sudden transverse pressure
  gradient at $t=0$. The profiles in blue and red represent the
  streamwise and spanwise mean velocity profiles, respectively. The
  channel flow is driven by a streamwise $\mathrm{d}P/\mathrm{d}x_1$
  and spanwise $\mathrm{d}P/\mathrm{d}x_3$ mean pressure gradient
  applied in the streamwise and spanwise direction, respectively.
  \label{fig:setup} }
\end{center}
\end{figure}

We perform a series of DNS of incompressible turbulent channel flow
subjected to a sudden imposition of a transverse pressure gradient
\citep{Moin1990}.  The problem set-up is sketched in figure
\ref{fig:setup}.  This flow configuration, yet simple, has proven
successful in capturing the essential features of non-equilibrium
3DTBL.  The calculation is initialised with a 2-D fully-developed
equilibrium channel flow.  At $t=0$, a mean spanwise pressure gradient
is applied, inducing a transient acceleration of the flow in the
spanwise direction. During this process, the channel flow is driven in
the streamwise direction by the usual mean streamwise pressure
gradient.  Our focus is on the initial transient succeeding the
application of the transverse pressure gradient.
%

\begin{figure}
\vspace{0.5cm}
\begin{center}
\includegraphics[width=1\textwidth]{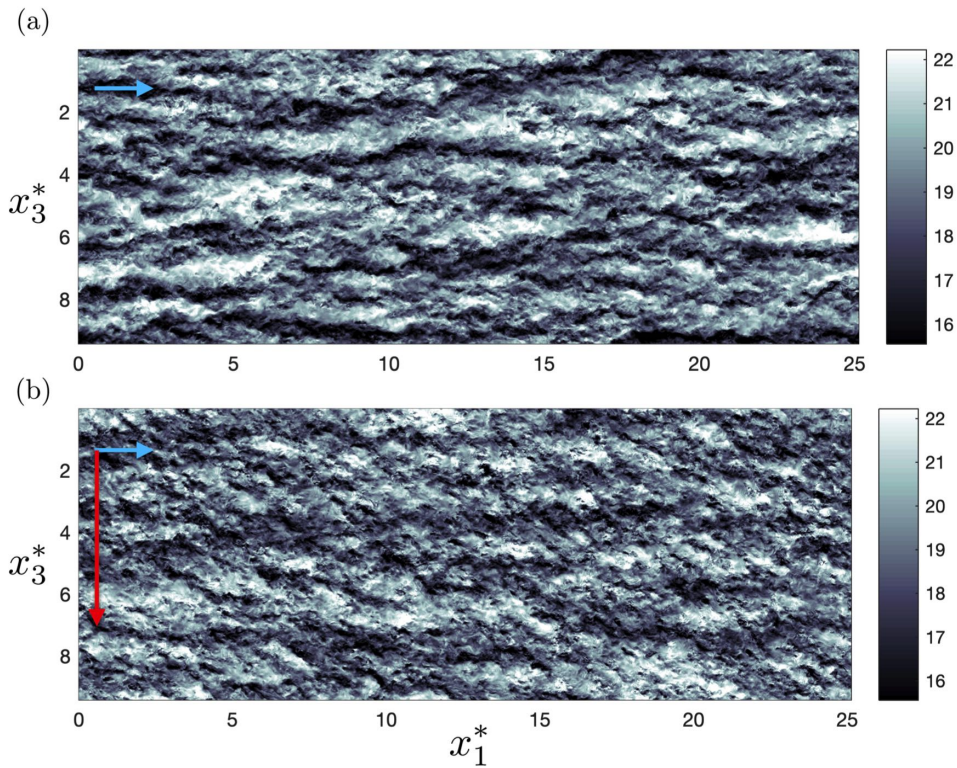}
\caption{ Instantaneous $x_1$--$x_3$ planes of the streamwise velocity
  at $x_2^* = 0.25$ for (a) $t^* = 0$, and (b) $t^* = 0.6$. The data
  is for $\Pi=60$ at $Re_\tau \approx 1000$. The colour bars show the
  magnitude of the streamwise velocity normalised in wall units. The
  arrows represent the direction of $\mathrm{d}P/\mathrm{d}x_1$ in
  blue and $\mathrm{d}P/\mathrm{d}x_3$ in red, but note that lengths
  of the arrows are not at scale.
  \label{fig:snapshots} }
\end{center}
\end{figure}

Two Reynolds numbers are considered, namely $Re_\tau=h
u_\tau/\nu\approx500$ and $Re_\tau\approx 1000$, both defined at
$t=0$, where $h$ is the channel half-height, $u_\tau$ is the friction
velocity at $t=0$, and $\nu$ is the kinematic viscosity.  The density
of the fluid is $\rho$. The streamwise, wall-normal, and spanwise
directions are represented by $x_1$, $x_2$, and $x_3$, respectively,
and the corresponding velocities are $u_1$, $u_2$, and $u_3$. The
pressure is denoted by $p$.  The size of the computational domain is
$L_1 \times L_2 \times L_3 = 4 \pi h \times 2h \times 2\pi h$ for
cases at $Re_\tau\approx500$, and $8 \pi h \times 2h \times 3\pi h$
for cases at $Re_\tau\approx 1000$. According to previous studies
\citep{Lozano2014a}, these domain sizes should suffice to accommodate
the largest structures populating the logarithmic layer $x_2 < 0.4h$
\citep{Marusic2013}.  Wall (or inner) units, $(\cdot)^+$, are obtained
by normalising flow quantities by $u_\tau$ and $\nu$, and outer units,
$(\cdot)^*$, are defined in terms of $u_\tau$ and $h$.  The streamwise
and spanwise mean pressure gradients are
$\mathrm{d}P/\mathrm{d}x_1=\rho u_\tau^2/h$ and
$\mathrm{d}P/\mathrm{d}x_3$, respectively. A campaign of simulations
at different $Re_\tau$ and multiple spanwise mean pressure gradients
are performed with spanwise to streamwise mean pressure gradient
ratios ranging from $\Pi = (\mathrm{d}P/\mathrm{d}x_3) /
(\mathrm{d}P/\mathrm{d}x_1) = 1,...,100$.  Several runs are considered
for each $Re_\tau$ and $\Pi$ by initialising the simulations with
various temporally-uncorrelated 2-D equilibrium turbulent channel
flows. The set of simulations is summarised in table
\ref{table1ch3}. Examples of the instantaneous streamwise velocity at
two time instants are shown in figure \ref{fig:snapshots} for $\Pi=60$
at $Re_\tau\approx 1000$.
%
\begin{table}
 \begin{center}
  \begin{tabular}{lcccccccccc} 
  \hline
    $Re_\tau$ & $L_1^*$ &  $L_3^*$ & $\Delta_1^+$ & $\Delta_3^+$ &  $\Delta_{2,\min}^+$ &  $\Delta_{2,\max}^+$ & $N_2$ & $T^*$ & $ \Pi $  & $N_R$ \\ \hline
        546 & $4\pi$ &  $2 \pi$ & 8.92 & 4.46 & 0.26 &  6.5  & $385$ & $1$ & 0,5,10,20,30,40,60,80 & 10 \\ 
        934 & $8\pi$ &  $3 \pi$ & 7.36 & 4.29 & 0.35 &  6.7  & $401$ & $1$ &  0,10,30,60,100 & 5 \\ 
  \hline
  \end{tabular}
  \caption{\label{table1ch3} Geometry and parameters of the DNS
    runs. $Re_\tau$ is the friction Reynolds number. $L_1^*=L_1/h$ and
    $L_3^*=L_3/h$ are the streamwise and spanwise dimensions of the
    numerical box, respectively, and $h$ is the channel half-height.
    $\Delta_1^+$ and $\Delta_3^+$ are the spatial grid resolutions in
    wall units for the streamwise and spanwise direction,
    respectively. $\Delta_{2,\min}^+$ and
    $\Delta_{2,\max}^+$ are the finer (closer to the wall) and
    coarser (further from the wall) grid resolutions in the
    wall-normal direction in wall units. $N_2$ is the number of
    wall-normal grid points. The simulations are integrated for a time
    $T^*$ equal to one eddy turnover, $T^*=Th/u_\tau=1$, where
    $u_\tau$ is the friction velocity. $\Pi =
    (\mathrm{d}P/\mathrm{d}x_3) / (\mathrm{d}P/\mathrm{d}x_1)$ is the
    spanwise to streamwise mean pressure gradient ratio driving the
    channel flow. $N_R$ is the total number of runs performed per each
    case given by the pair ($Re_\tau,\Pi$). }
 \end{center}
\end{table}
%

The simulations are performed by discretising the incompressible
Navier-Stokes equations with a staggered, second-order accurate,
centred, finite difference method \citep{Orlandi2000} in space, and a
explicit third-order accurate Runge-Kutta method \citep{Wray1990} for
time advancement. The system of equations is solved via an operator
splitting approach \citep{Chorin1968}.  Periodic boundary conditions
are imposed in the streamwise and spanwise directions, and the no-slip
condition is applied at the walls. The code has been validated in
turbulent channel flows \citep{Lozano2016_brief,Bae2018a} and
flat-plate boundary layers \citep{Lozano2018a}. The streamwise and
spanwise grid resolutions are uniform and denoted by $\Delta_1$ and
$\Delta_3$, respectively. The wall-normal grid resolution, $\Delta_2$,
is stretched in the wall-normal direction following an hyperbolic
tangent. The time step is such that the Courant-Friedrichs-Lewy
condition is always below 0.5 during the run. Details on the
parameters of the numerical set-up are included in table
\ref{table1ch3}.

\section{Analysis of non-equilibrium 3DTBL}\label{sec:analysis}


The present section is devoted to, first, the identification of
universal scaling laws for the tangential Reynolds stress in the 3-D
transient channel flow described in \S\ref{sec:numerics}, and second,
the scrutiny of the structural and energetic alterations of the flow
during the transient. A large number of studies have been dedicated to
the scaling of quantities of interest in fully-developed 2DTBL
\citep[see e.g.,][]{Millikan1938, Klewicki2007,Monkewitz2008}. Recent
efforts have been facilitated by the increased availability of
numerical data at high Reynolds numbers with an appreciable scale
separation between the inner and outer layers. On the contrary,
advances in non-equilibrium 3DTBL have been hindered by the lack of
high Reynolds number flow datasets. Similar limitations apply to the
analysis of structural changes on the flow.

The next section offers an overview of the time evolution of the
one-point statistics during the transient period, followed by a
discussion on the role of the no-slip wall. Then, we classify the flow
regimes and analyse the scaling laws concerning the time history of
the tangential Reynolds stress. The time-dependent, 3-D structural
changes undergone by the flow are discussed at the end of the section,
where we propose a structural model consistent with our observations.

\subsection{Overview of one-point statistics}\label{sec:overview}

We select the channel flow at $Re_\tau \approx 500$ with $\Pi = 60$ as
a representative case to illustrate the non-equilibrium response of
the flow succeeding the imposition of the lateral pressure
gradient. The systematic analysis for various $Re_\tau$ and $\Pi$ is
presented in \S\ref{subsec:regimes}.  For $t\rightarrow \infty$, the
system attains an new statistically steady state corresponding to a
2-D channel flow at higher $Re_\tau$ and mean-flow direction parallel
to the vector
$(\mathrm{d}P/\mathrm{d}x_1,0,\mathrm{d}P/\mathrm{d}x_3)$. We focus on
the initial transient dominated by 3-D non-equilibrium effects for
$t^*<1$. The statistical quantities of interest are computed by
averaging the flow in the homogeneous directions, over the top and
bottom halves of the channel, and among different runs. The averaging
operator is hereafter denoted by $\langle \cdot \rangle$, and velocity
fluctuations are signified by $(\cdot)'$.  Fluctuating velocities are
measured with respect to the time-evolving mean velocity profiles in
the streamwise and spanwise direction, $\langle u_1 \rangle(x_2,t)$
and $\langle u_3 \rangle(x_2,t)$, respectively.

\begin{figure}
\vspace{0.5cm}
\begin{center}
\includegraphics[width=1\textwidth]{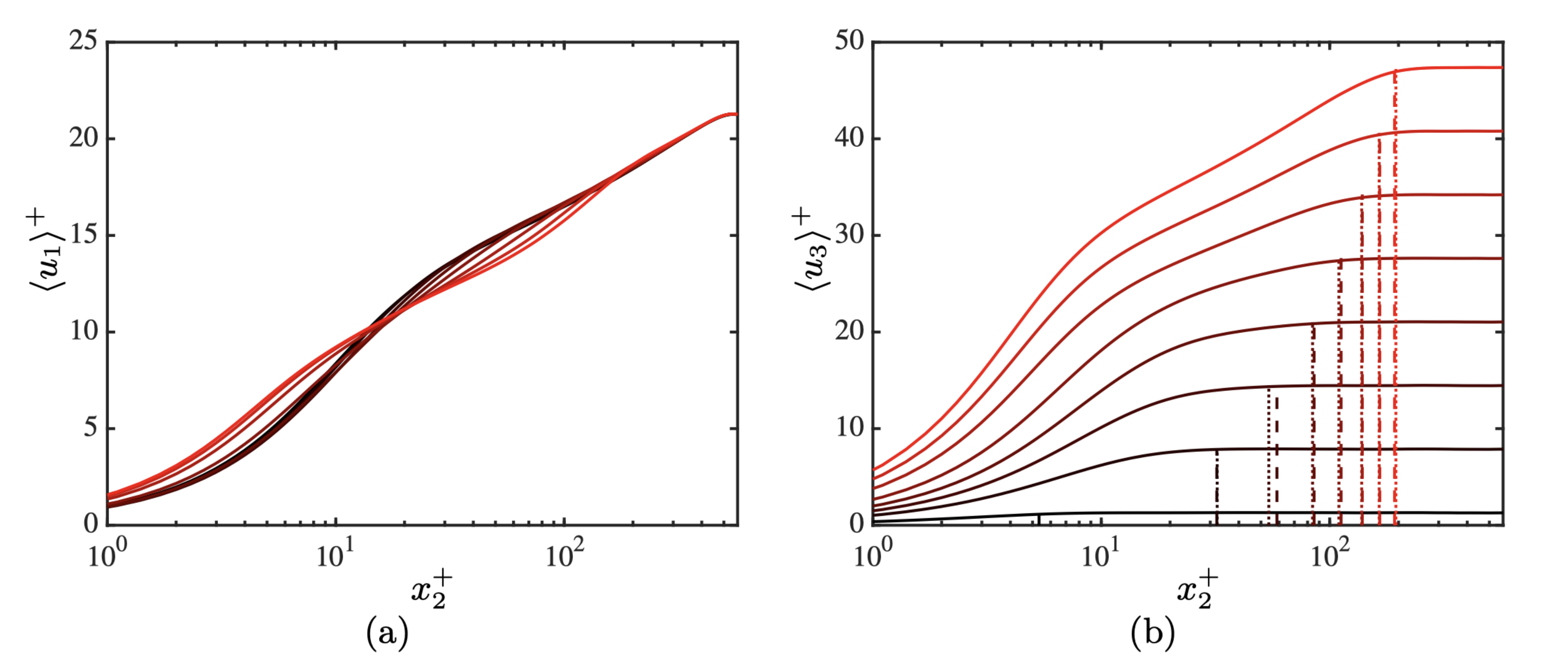}
 \end{center}
\caption{ Mean velocity profile in (a) the streamwise direction and
  (b) the spanwise direction for $t^+=12, 72, 132, 192, 252, 312,
  372$, and $432$. Colours indicate time from $t^+=0$, black, to
  $t^+=432$, red.  The vertical dotted lines (\dotted) are the
  boundary layer thickness $\delta_3$ defined by the wall-normal
  distance at which $\langle u_3 \rangle = 0.99\langle u_3
  \rangle_\infty=0.99\langle u_3 \rangle(h,t)$, and the vertical
  dashed lines (\dashed) are the estimated boundary layer thickness
  given by $\delta_3^+ = 0.445 t^+$. \label{fig:means} }
\end{figure}

The mean velocity profiles are shown in figure \ref{fig:means} at
several time instants.  The streamwise mean velocity undergoes mild
changes in shape (figure \ref{fig:means}a), and the main outcome of
the lateral pressure gradient is the development of a spanwise
boundary layer of thickness $\delta_3$ (figure \ref{fig:means}b).  The
growth of $\delta_3$ is initially governed by viscous diffusion, i.e.,
$\delta_3 \sim \sqrt{\nu t}$ for $t<t_\nu$. A rough estimation of
$t_\nu$ is given by $t_\nu^+ \approx 50$ \citep{Moin1990}, and the
initial viscous growth can be neglected at high $Re_\tau$. For
$t>t_\nu$, turbulent diffusion prevails and $\delta_3 \sim \sqrt{
  \nu_e t}$, where $\nu_e$ is the turbulent eddy-viscosity. Assuming
the mixing-length hypothesis, $\nu_e \sim u_\tau \delta_3$, then
$\delta_3 \sim u_\tau t$, i.e. the spanwise boundary layer grows
linearly in time regardless of $\mathrm{d}P/\mathrm{d}x_3$ in first
order approximation. The prediction of $\delta_3^+ \approx 0.445 t^+$,
included in figure \ref{fig:means}(b), highlights the validity of the
previous assumptions after the initial viscous phase. The inertial
core of the channel, $\langle \cdot \rangle_\infty$, is accelerated by
the mean spanwise pressure gradient such that $\rho \langle u_3
\rangle_\infty \approx \mathrm{d}P/\mathrm{d}x_3 t$, which controls
the additional spanwise shear, $\partial \langle u_3 \rangle /\partial
x_2 \sim \langle u_3 \rangle_\infty/\delta_3 \sim
\mathrm{d}P/\mathrm{d}x_3 / (\rho u_\tau)$. In summary, the sudden
imposition of $\mathrm{d}P/\mathrm{d}x_3$ results in the emergence of
a spanwise boundary layer diffusing upwards the wall linearly in time,
$\delta_3 \sim u_\tau t$, accompanied by an additional mean shear
proportional to $\mathrm{d}P/\mathrm{d}x_3$.
%

\begin{figure}
\vspace{0.5cm}
%
%
\begin{center}
\includegraphics[width=1.0\textwidth]{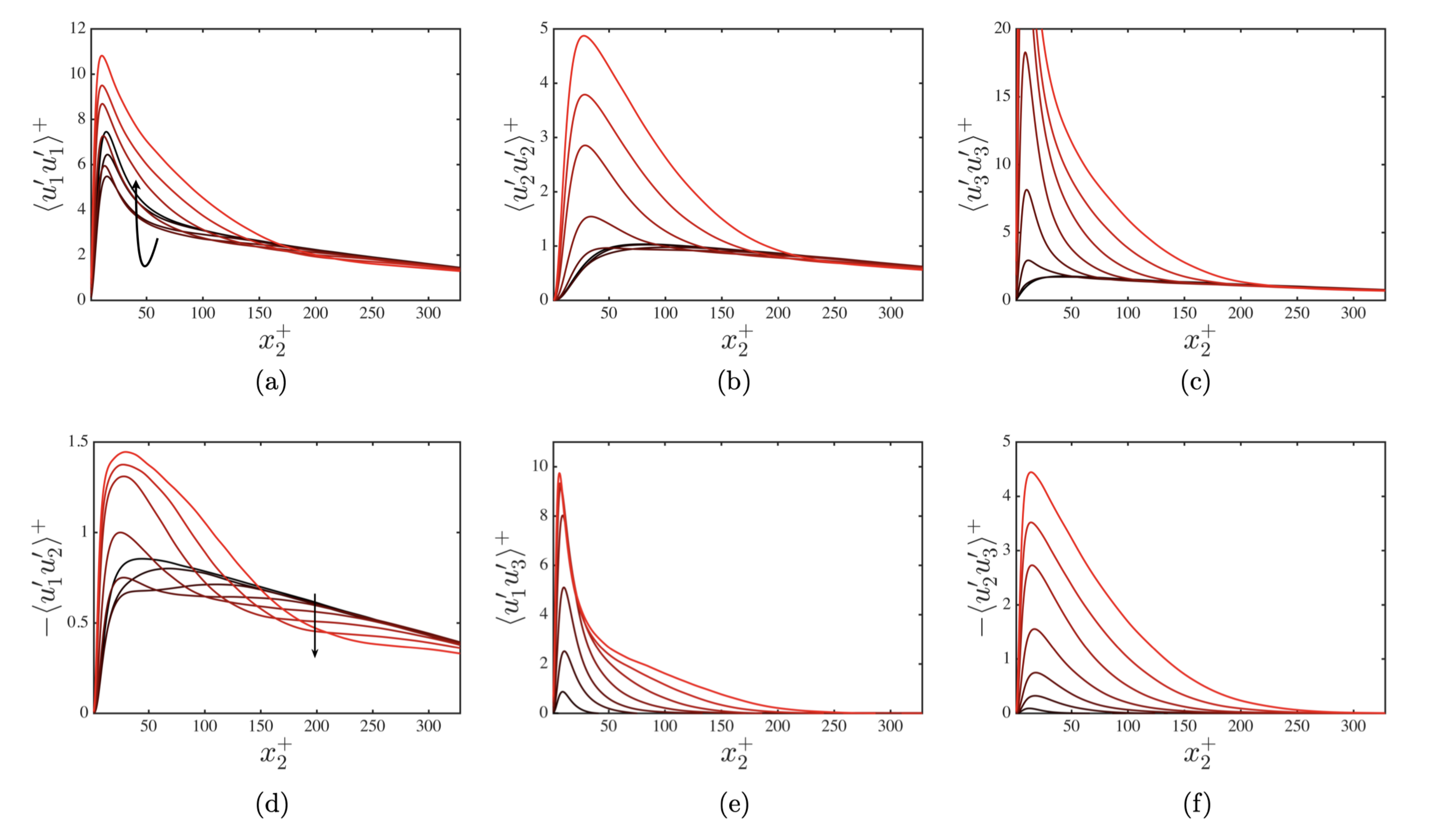}
\end{center}
\caption{ Mean Reynolds stress for $\Pi=60$ at $Re_\tau \approx 500$.
  Different lines correspond to different times at $t^+=12, 72, 132,
  192, 252, 312, 372$, and $432$. Colours indicate time from $t^+=0$,
  black, to $t^+=432$, red. The arrows indicate the direction of time.
  \label{fig:stresses} }
\end{figure}

\begin{figure}
\vspace{0.5cm}
\begin{center}
   \vspace{0.1cm}
\includegraphics[width=1\textwidth]{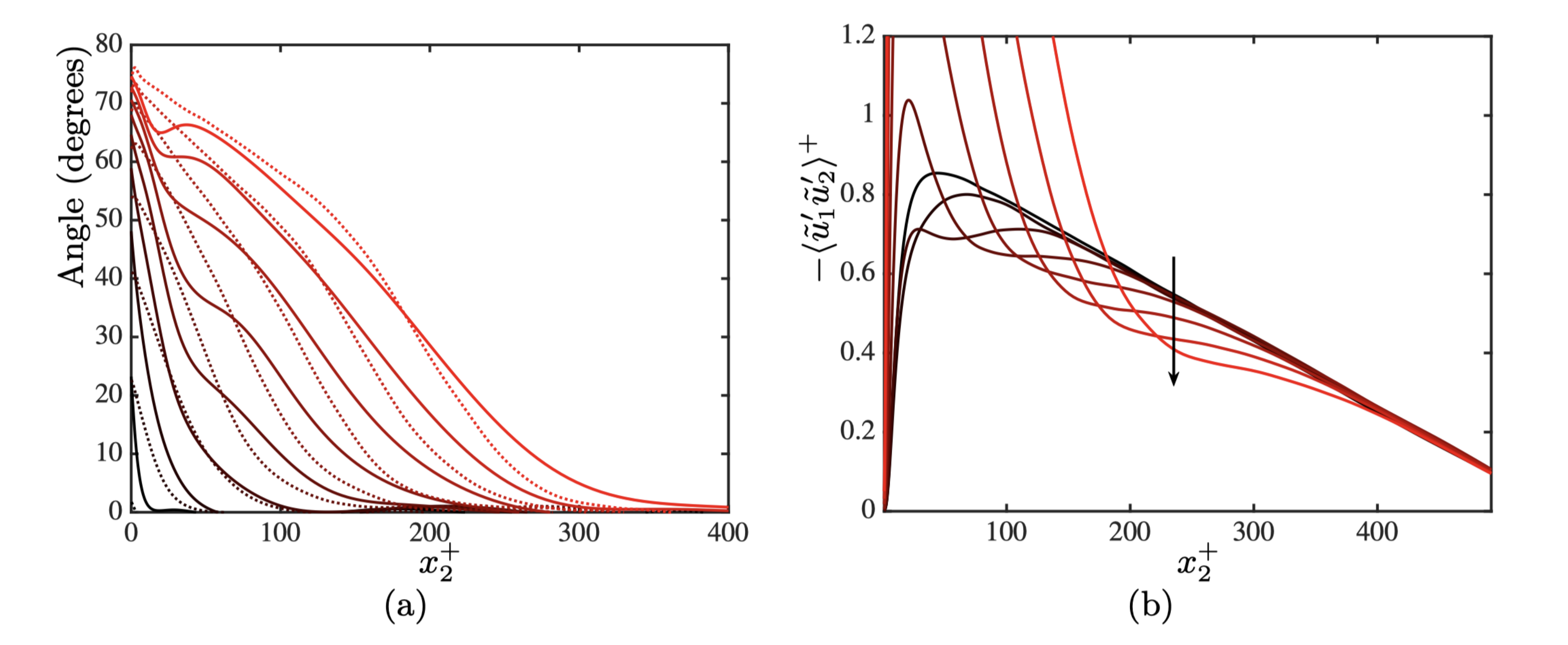}
 \end{center}
\caption{ (a) Angle of the mean Reynolds stress direction
  $\gamma_\tau$ (\solid) and mean shear direction $\gamma_S$ (\dotted)
  with respect to $x_1$. (b) Mean tangential Reynolds stress in the
  wall-normal and time-dependent frame of reference
  $\tilde{\mathcal{F}}$ aligned with the mean-shear direction
  $\gamma_S$. The arrow in panel (b) indicates the direction of time.
  Different lines correspond to different times at $t^+=12, 72, 132,
  192, 252, 312, 372$, and $432$. In both panels, the colours denote
  time from $t^+=0$, black, to $t^+=432$, red.  The data is for
  $\Pi=60$ at $Re_\tau \approx 500$.
  \label{fig:stress_aligned} }
\end{figure}

The time evolution of the mean Reynolds stresses is shown in figure
\ref{fig:stresses}.  Considering that the flow is subjected to the
additional strain $\partial \langle u_3 \rangle/\partial x_2$, the
classic theory anticipates an increase of the Reynolds stresses under
the equilibrium assumption $-\langle u'_i u'_j \rangle + 1/3 \langle
u_k' u_k'\rangle\delta_{ij} \propto \nu_e \langle S_{ij} \rangle$,
where $S_{ij}$ is the rate-of-strain tensor and $\delta_{ij}$ is the
Kronecker delta. Figure \ref{fig:stresses} shows that the behaviour of
$\langle u_i' u_j'\rangle$ is consistent with the equilibrium
prediction for long times. However, $\langle u_1'u_1'\rangle$ and
$-\langle u_1' u_2'\rangle$ experience a vigorous depletion during the
first stages of the transient, whereas the other stresses remain
roughly constant inconsistent with the equilibrium assumption. The
reduction in magnitude of those stresses comprising $u_1'$ hints to a
deficiency in the streak generation cycle triggered during the
transient, and the structural origin of such a deficiency is discussed
in \S\ref{subsec:structure}.  A similar equilibrium argument applies
to the angle of Reynolds stress direction, $\gamma_\tau=
\mathrm{atan}[\langle u_2'u_3' \rangle/ \langle u_1'u_2' \rangle]$,
and mean shear direction $\gamma_S = \mathrm{atan}[ (\partial \langle
  u_3 \rangle / \partial x_2)/(\partial \langle u_1 \rangle / \partial
  x_2) ]$, which are expected to satisfy $\gamma_\tau \approx
\gamma_S$ in equilibrium 2DTBL. As seen from figure
\ref{fig:stress_aligned}(a), the equilibrium condition is not met for
the angles; the Reynolds stress direction lags behind the mean
direction closer to the wall and leads further away.  We will focus
most of our attention on the tangential Reynolds stress, $-\langle
u_1'u_2' \rangle$, because the initial non-equilibrium response is
most vividly manifested on that component, although other metrics can
be defined to measure non-equilibrium effects such as the classic
Townsend's structure parameter \citet{Townsend1976}.
%

It could be argued that the drop in $-\langle u_1' u_2' \rangle$ in
figure \ref{fig:stresses}(d) is an artefact of the static frame of
reference $\mathcal{F}:(x_1,x_2,x_3)$. The direction given by
$\mathcal{F}$ is no longer co-planar to the mean shear vector, which
is the primal source responsible for the injection of kinetic energy
into the turbulence intensities. To show that the depletion of
$-\langle u_1'u_2' \rangle$ is not the consequence of observing the
flow from the point of view of $\mathcal{F}$, we define the
wall-normal and time-dependent frame of reference
$\tilde{\mathcal{F}}:(\tilde x_1,x_2,\tilde x_3)$ such that $\tilde
x_1$ points in the direction of the local mean shear vector $(\partial
\langle u_1 \rangle/\partial x_2, 0 ,\partial \langle u_3
\rangle/\partial x_2)$ at each wall-normal location and time
instant. The angle between $\tilde x_1$ and $x_1$ is given by
$\gamma_S$ (figure \ref{fig:stress_aligned}a). The velocity components
in the frame of reference $\tilde{\mathcal{F}}$ are denoted by $\tilde
u_1$, $\tilde u_2$$(\equiv$$u_2)$, and $\tilde u_3$. Figure
\ref{fig:stress_aligned}(b) demonstrates that the shear-aligned
tangential Reynolds stress, $-\langle \tilde u_1' \tilde u_2'
\rangle$, also experiences a strong reduction in magnitude. An
alternative frame of reference is that aligned with the principal
Reynolds stress direction defined by the angle $\gamma_\tau$
\citep{Moin1991}. The difference between $\gamma_S$ and $\gamma_\tau$
is small (figure \ref{fig:stress_aligned}a), and the time history of
the Reynolds stresses in the frame of reference of the principal
Reynolds stress direction (not shown) is comparatively similar to the
results from figure \ref{fig:stress_aligned}(b).
%

\subsection{Reynolds stress depletion due to uniform acceleration 
versus lateral boundary layer growth}\label{sec:dw0}

The drop in $-\langle u_1' u_2'\rangle$ occurs in concomitance with
the growth of the spanwise shear layer.  Thus, it was presupposed in
the analysis above that the deficit in $-\langle u_1' u_2' \rangle$
originates from the wall and spreads toward the outer layer at the
rate dictated by $\mathrm{d} \delta_3/\mathrm{d}t$. In this section,
we assess whether the Reynolds stress reduction is caused by
aforementioned emergence of a strong shear layer or, on the contrary,
by the spanwise uniform acceleration of the flow. The discussion is
relevant as accelerating flows are a basic resource of experimental
facilities (e.g., contractions in wind tunnels), aiming to reduce the
turbulence intensity levels and anisotropy
\citep[see][pp. 68]{Batchelor1953}.

\begin{figure}
\vspace{0.5cm}
\begin{center}
   \vspace{0.1cm}
\includegraphics[width=1\textwidth]{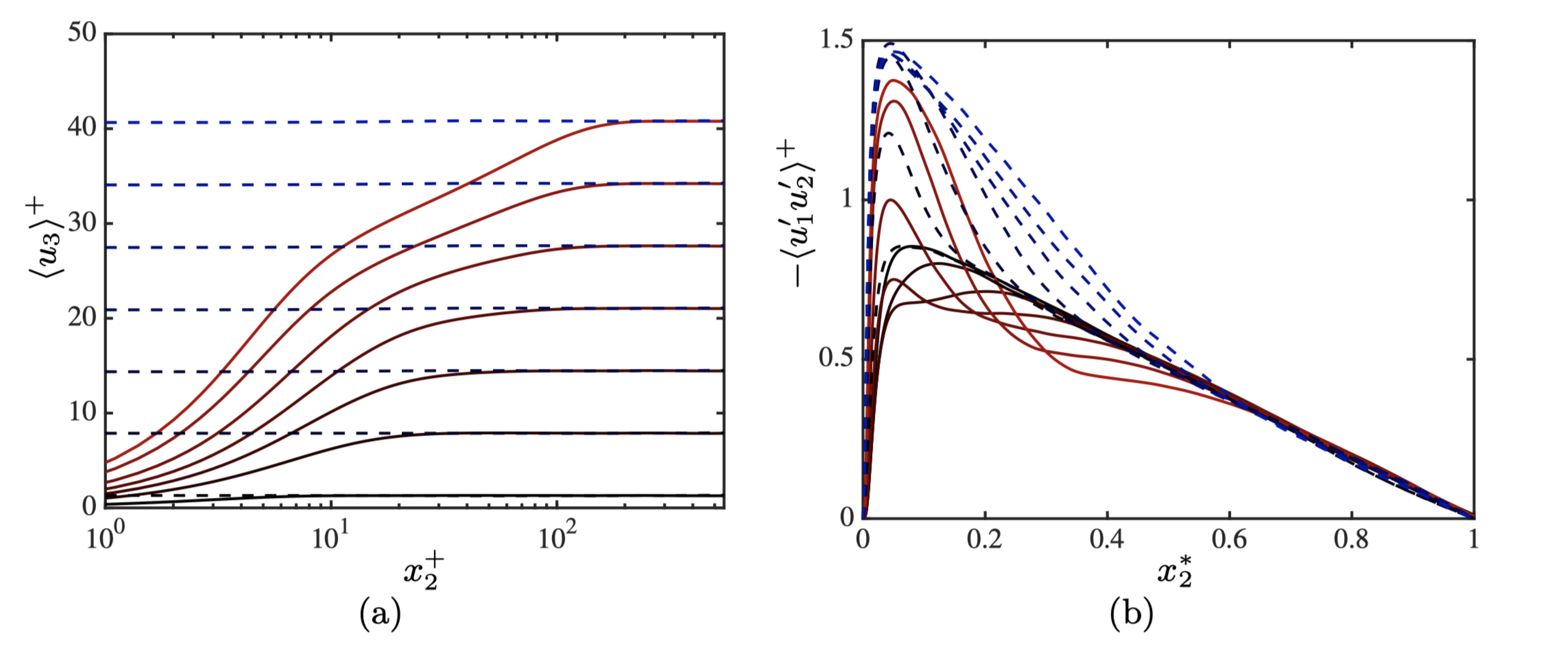}
 \end{center}
\caption{ (a) Mean spanwise velocity profile and (b) mean tangential
  Reynolds stress for $\Pi= 60$ at $Re_\tau\approx 500$. The different
  lines correspond to $t^+=12, 72, 132, 192, 252, 312, 372$, and
  $432$. Lines (\solid, reddish) are for a channel with no-slip wall,
  and lines (\dashed, bluish) are for a channel with free-slip
  boundary condition in the spanwise direction. The colours indicate
  time from $t^+=0$, black, to $t^+=432$, red or blue.
  \label{fig:dwd0} }
\end{figure}

To isolate the effect of the lateral boundary layer from the spanwise
acceleration, we conduct a DNS channel flow in a similar set-up to
\S\ref{sec:numerics} but enforcing free-slip boundary conditions for
$u_3$ instead of no-slip walls. The case considered is for $\Pi = 60$ at
$Re_\tau\approx500$. The time evolution of the mean
spanwise velocity and tangential Reynolds stress under such settings
are reported in figure \ref{fig:dwd0} and compared with the analogous
no-slip case. The free-slip in $u_3$ allows for an accelerating plug
flow in $x_3$, in which wall-blocking effects are still present but
the formation of a spanwise shear layer is inhibited (figure
\ref{fig:dwd0}a). As is apparent from figure \ref{fig:dwd0}(b), the
spanwise acceleration alone does not entail a reduction of $-\langle
u_1' u_2'\rangle$. Hence, we conclude that the spanwise shear layer
developing from the wall must be regarded as the source of the
counter-intuitive drop of tangential Reynolds stress.
%

\subsection{Flow regimes}\label{subsec:regimes}

We quantify the flow regimes of the transient response of the
momentum-carrying eddies (responsible for $-\langle u_1' u_2'\rangle$)
subjected to non-equilibrium effects.  We can anticipate that for low
values of $\Pi$, the perturbation introduced by the lateral forcing is
very gentle and eddies evolve in a quasi-equilibrium state
irrespective of their size and lifespan. Conversely, large values of
$\Pi$ are expected to drive the entire population of eddies at all
scales across the boundary out of equilibrium. The non-dimensional
parameters governing these flow regimes are $Re_\tau$ and $\Pi$.

The level of non-equilibrium endured by the momentum-carrying eddies
can be estimated by assuming that, prior to the application of $\Pi$,
the boundary layer is populated by a collection of wall-attached
self-similar eddies with sizes $l_e$ proportional to the distance to
the wall, $l_e\sim x_2$, and characteristic velocity $u_\tau$
\citep{Townsend1976}. Consistently, the characteristic lifetime of
eddies of size $l_e$ is $t_e\sim x_2 / u_\tau$. The smallest
momentum-carrying eddies are found close to the wall at $x_2 \sim
\nu/u_\tau$ due to the limiting effect of viscosity, and their
lifetimes reduce to $t_e \sim \nu/u_\tau^2$.  The largest eddies are
constrained by the channel height $x_2\sim h$, with lifetimes $t_e
\sim h/u_\tau$. The lateral mean pressure gradient introduces an
additional time-scale associated with the spanwise acceleration of the
flow $t_p \sim \rho u_\tau / (\mathrm{d}P/\mathrm{d}x_3)$.  The
condition for non-equilibrium is $t_p < t_e$, i.e. the characteristic
time to accelerate the flow in the spanwise direction is shorter than
the lifetime of the momentum-carrying eddies in order to shove the
latter out of the equilibrium state. A similar conclusion is drawn
reasoning in terms of the minimum strength of the lateral shear layer
$\sim$$\mathrm{d}P/\mathrm{d}x_3 / (\rho u_\tau)$ necessary to disturb
the local shear of the wall-attached eddies
$\sim$$u_\tau/x_2$. 

\begin{figure}
\vspace{0.5cm}
\begin{center}
\includegraphics[width=1\textwidth]{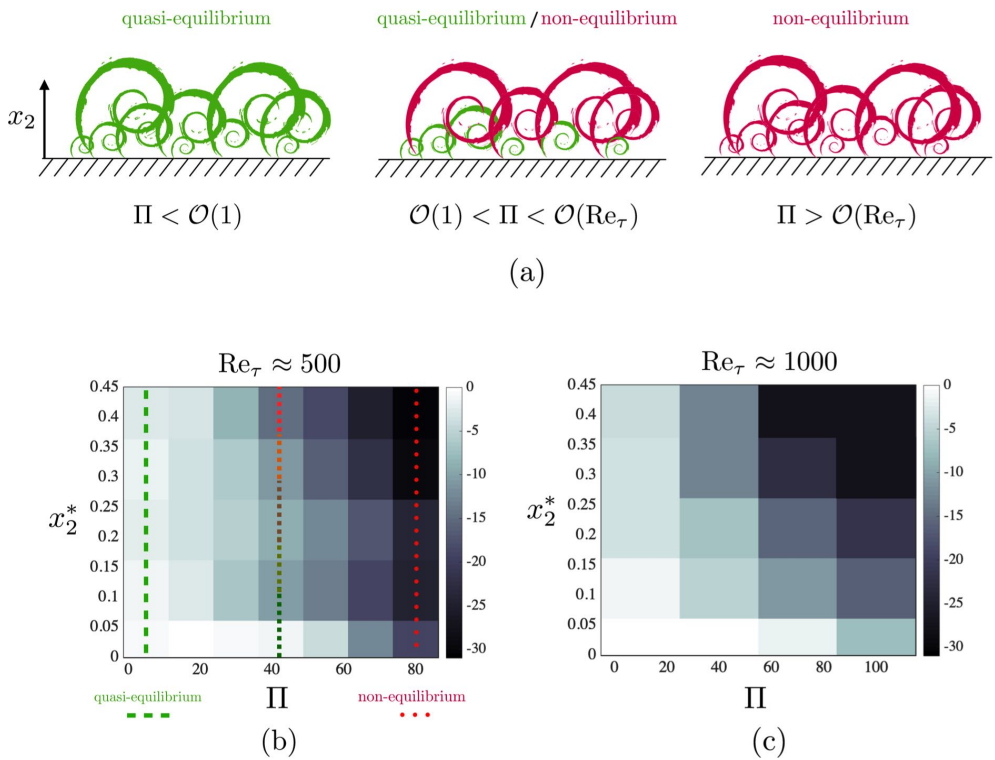}
\caption{ (a) Schematic of self-similar, wall-attached,
  momentum-carrying eddies, and different flow regimes as function of
  the spanwise to streamwise mean pressure gradient ratio $\Pi$. The
  eddies coloured in green are in a quasi-equilibrium state, whereas
  eddies coloured in red are out-of-equilibrium. Panels (b) and (c)
  are the percentage drop of tangential Reynolds stress, $\min_{t} \{
  D_\tau \}$, in the frame of reference of the mean shear
  $\tilde{\mathcal{F}}$ as a function of the spanwise to streamwise
  the mean pressure gradient ratio $\Pi$ and wall-normal distance
  $x_2^*$ for (b) $Re_\tau\approx 500$ and (c) $Re_\tau\approx
  1000$. The vertical lines in (b) represent flow states ranging from
  the equilibrium regime (green) to non-equilibrium regime (red).
  \label{fig:regimes} }
\end{center}
\end{figure}

Based on the flow scales discussed above, we differentiate three flow
regimes as sketched in figure \ref{fig:regimes}(a). For $\Pi <
\mathcal{O}(1)$ ($t_e > t_p$), the spanwise pressure gradient is
categorised as weak, and all flow scales relax instantly to a
quasi-equilibrium state during the transient period. Conversely, for
$\Pi > \mathcal{O}(Re_\tau)$ ($t_e < t_p$), the momentum-carrying
eddies are unable to adjust to the prompt imposition of the shear
regardless of their size. For intermediate values of $\Pi$, eddies
coexist in both quasi-equilibrium and non-equilibrium states, the
former being the eddies located in the region closer to the wall.
%

The analysis above is corroborated in figures \ref{fig:regimes}(b) and
(c), which show the maximum percentage drop of the tangential Reynolds
stress during the transient period after the imposition of the lateral
mean pressure gradient, $\min_{t} \{ D_\tau(x_2,t) \}$, where $D_\tau$
is defined as
\begin{equation}
 D_\tau(x_2,t) = \frac{ \langle \tilde u_1 \tilde u_2 \rangle(x_2,t) -\langle \tilde u_1 \tilde u_2 \rangle(x_2,0)  }
{ \langle \tilde u_1 \tilde u_2 \rangle(x_2,0) } \times 100.
\label{eq:Dtau}
\end{equation}
Note that the Reynolds stress in (\ref{eq:Dtau}) is referred to the
frame of reference $\tilde{\mathcal{F}}$ aligned with the mean
shear. Similar conclusions are drawn when the stress is referred to
$\mathcal{F}$. The results in figure \ref{fig:regimes}(b) reveal that
the relative reduction in the Reynolds stress attains up to 30\%, and
that the drop accentuates for increasing $\Pi$ and $x_2^*$.  Figure
\ref{fig:regimes}(c) confirms that the trend holds at higher
$Re_\tau$.
%
\begin{figure}
\vspace{0.5cm}
\begin{center}
   \vspace{0.1cm}
\includegraphics[width=1\textwidth]{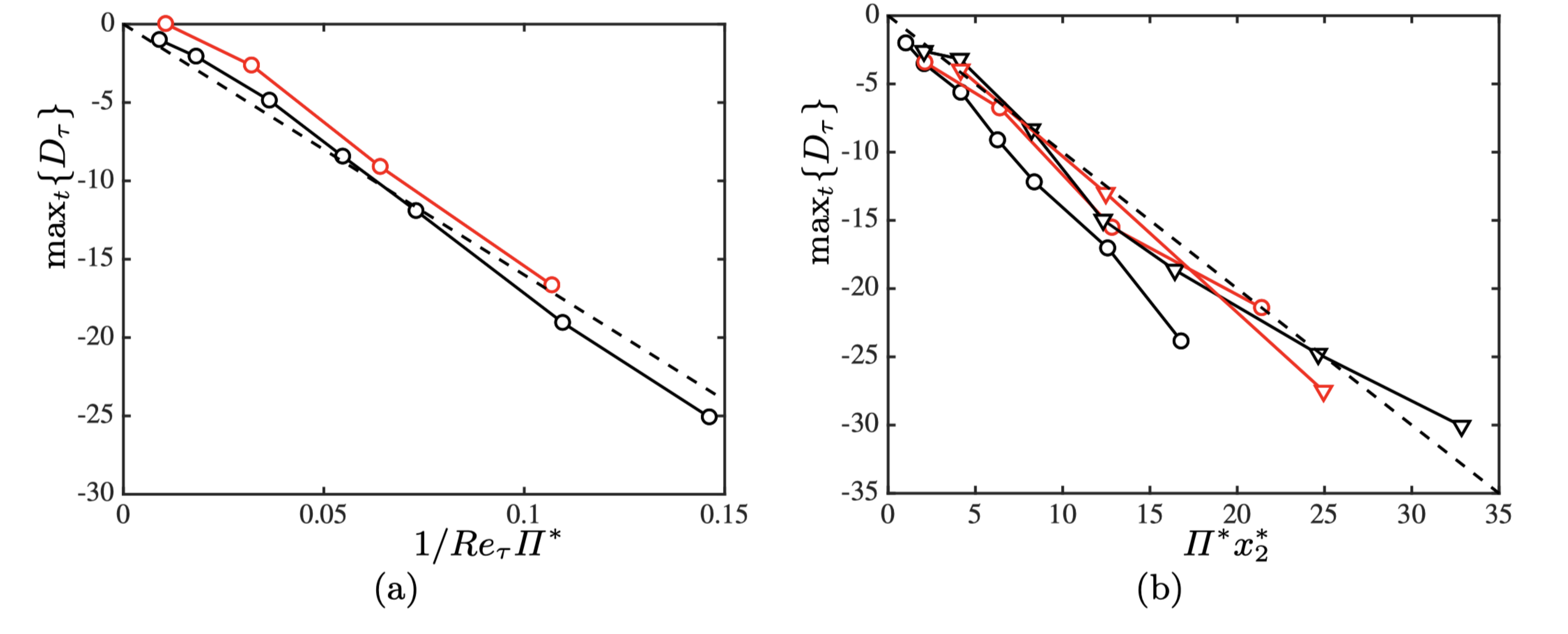}
 \end{center}
\caption{ Maximum percentage drop of the tangential Reynolds stress
  $\min_{t} \{ D_\tau \}$ in the frame of reference of the mean
  shear. Colours are black for cases at $Re_\tau\approx 500$ and red
  for cases at $Re_\tau\approx 1000$. In panel (a), lines with
  (\circle) are $\min_{t} \{ D_\tau \}$ at $x_2^+ =30$, and (\dashed)
  is $\min_{t} \{ D_\tau \} \approx -160 \Pi/Re_\tau$.  In
  panel (b), lines with symbols are $\min_{t} \{ D_\tau \}$ at
  $x_2^*=0.2$ (\circle) and $x_2^*=0.4$ ($\triangledown$), and
  (\dashed) corresponds to $\min_{t} \{ D_\tau \} \approx -\Pi x_2^*$.
  \label{fig:dP_drop} }
\end{figure}

The scaling of $\min_t\{ D_\tau \}$ is inspected in figure
\ref{fig:dP_drop}, which contains various cuts of the
$(\Pi,x_2^*)$-maps shown in figures \ref{fig:regimes}(b) and
(c). Within the buffer region (figure \ref{fig:dP_drop}a), the
response of the flow is controlled by the viscous scales.  The
momentum equation in inner units is given by
\begin{equation}
	\frac{\mathcal{D} u_i^+}{\mathcal{D}t^+} = -\frac{\partial p'^+}{\partial x_i^+} +
        \frac{\mathrm{d}P^+}{\mathrm{d} x_i^+}\delta_{i3} +
        \frac{\partial ^2 u_i^+}{\partial x_k^+ \partial x_k^+} \, ,
	\label{momentum_eq_inner}
\end{equation}
where $\mathcal{D}$ denotes material derivative,
$\mathrm{d}P^+/\mathrm{d}x_1^+=\mathcal{O}(1/Re_\tau)$ has been
neglected, and summation through the repeated index $i$ is not meant
for the third term on the right-hand side. From
(\ref{momentum_eq_inner}), we conclude that a similar reduction in the
Reynolds stress is obtained across different $Re_\tau$ for identical
values of $\Pi /Re_\tau$, which is the relevant spanwise to streamwise
mean pressure gradient for the buffer region.

For the logarithmic layer thus at high $Re_\tau$ (figure
\ref{fig:dP_drop}b), wall-attached eddies of a given size $l_e\sim
x_2$ experience a similar drop in the Reynolds stress when the mean
spanwise pressure gradient is normalised by the characteristic scales,
$x_2$ and $u_\tau$, controlling the eddies.  Analysis of the
nondimensional equations obtained by introducing the similarity
variable $\eta = t/t_e = t u_\tau/x_2$ reveals that the condition for
self-similar Reynolds stress depletion at a given wall-normal distance
is obtained by a common value of the compensated spanwise to
streamwise mean pressure gradient ratio, $\Pi x_2^*$, consistent with
the results from figure \ref{fig:dP_drop}(b).
%

From the scaling analysis above and the numerical results in figure
\ref{fig:dP_drop}, the quantitative drop in Reynolds stress for the
flow motions free of viscous effects at a given $x_2$ location is well
approximated by
\begin{equation}
\min_t\{D_\tau\} \approx -\Pi x_2^*.
	\label{Dtau_log}
\end{equation}
If we further assume that the self-similar scaling of the flow motions
with $x_2$ does not hold below $x_2^+ \approx 160$, the inner layer
scaling law for the Reynolds stress drop from (\ref{Dtau_log}) reduces
to
\begin{equation}
\min_t\{D_\tau\} \approx -160 \frac{\Pi}{Re_\tau},
	\label{Dtau_inner}
\end{equation}
which is valid for the buffer region and serves as an approximation to
the trends observed in figure \ref{fig:dP_drop}(a).

Finally, a tentative relation delimiting the necessary spanwise
forcing to achieve the fully non-equilibrium regime (eddies out of
equilibrium across the entire  boundary layer), arbitrarily delimited
by $\min_t\{ D_\tau \} < -5\%$, is given by
\begin{equation}
\Pi > 0.03 Re_\tau.
\label{eq:regimes}
\end{equation}
Equation (\ref{eq:regimes}) shows that the lateral mean pressure
gradient required to attain the fully non-equilibrium regime increases
proportionally to the Reynolds number. The meaning of $\Pi $ in this
particular flow cannot be unambiguously extrapolated to more general
flows configurations. Nonetheless, the time-scale argument used to
derived (\ref{eq:regimes}) suggest that, in external aerodynamic
applications, the inner layer is most likely to be found in a
quasi-equilibrium state given the high Reynolds numbers typically
encountered in these situations.

\subsection{Time evolution of the tangential Reynolds stress}

In the previous section we were concerned with the maximum drop in the
tangential Reynolds stress without consideration of its time
response. Here, we discuss the scaling of the time evolution of
$D_\tau$ for 3-D channels in the fully non-equilibrium regime, i.e.,
$\Pi > 0.03 Re_\tau$, which is the most intriguing case from the
physical viewpoint. As in \S\ref{subsec:regimes}, we perform the
analysis separately for the buffer region and logarithmic layer,
although the former can be thought of as the near-wall limit of the
latter.

\begin{figure}
\vspace{0.5cm}
\begin{center}
   \vspace{0.1cm}
 \includegraphics[width=1\textwidth]{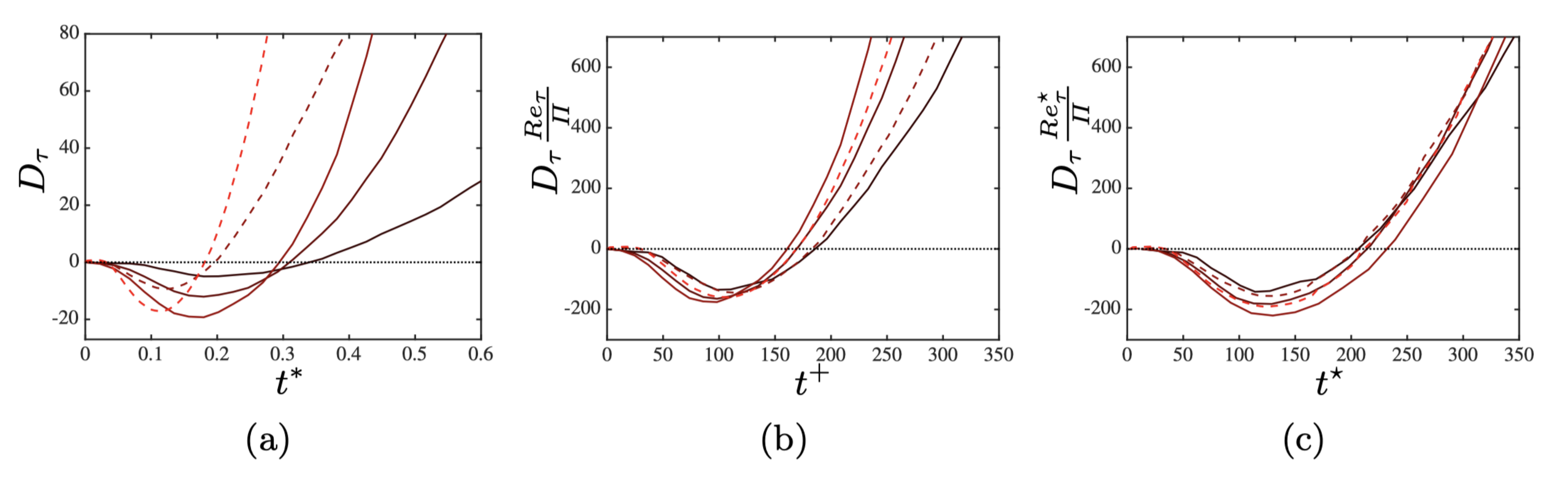}
 \end{center}
\caption{ Time evolution of the percentage change of tangential
  Reynolds stress $D_\tau$ in the buffer layer for $x_2^+=30$ in
  panels (a) and (b), and for $x_2^\star=30$ in panel (c). The lines
  are (\solid) for $Re_\tau \approx 500$ and (\dashed) for $Re_\tau
  \approx 1000$. For cases at $Re_\tau \approx 500$, colours are
  $\Pi=20,40$, and $60$ from dark red to light red. For cases at
  $Re_\tau \approx 1000$, colours are dark red for $\Pi=60$ and light
  red for $\Pi=100$.
  \label{fig:scaling_inner} }
\end{figure}

The time evolution of $D_\tau$ in the buffer layer is plotted in
figure \ref{fig:scaling_inner} for various pairs of
$(Re_\tau,\Pi)$. Three scalings are inspected. Figure
\ref{fig:scaling_inner}(a) shows the evolution of $D_\tau$ as a
function of time normalised in outer units. Unsurprisingly, both the
intensity of $D_\tau$ and the time instant for the maximum drop varies
considerably among distinct combinations of $(Re_\tau,\Pi)$. Inasmuch
as the near-wall eddies do not scale in outer units, the results in
figure \ref{fig:scaling_inner}(a) are included only to expose the lack
of collapse among cases under an inadequate normalisation. The
time-scaling using wall units is tested in figure
\ref{fig:scaling_inner}(b).  It was argued in \S\ref{subsec:regimes}
that the depletion of Reynolds stress within the inner layer is
proportional to $\Pi /Re_\tau$. Consistently, the results in figure
\ref{fig:scaling_inner}(b) are plotted against the compensated
Reynolds stress drop, $D_\tau Re_\tau/\Pi$. The new scaling improves
the collapse of the results, especially for $t^+<150$, above which the
time evolution of $D_\tau Re_\tau/\Pi$ diverges among cases. The
absence of collapse for $t^+>150$ coincides with the typical lifetime
of the momentum-carrying eddies in the buffer layer
\citep{Lozano2014b}. Thus, $u_\tau$ (defined at $t=0$) is
representative of the originally-in-equilibrium near-wall eddies until
the generation cycle is restarted and newborn eddies emerge under
different flow conditions. Following the previous reasoning, the
collapse can be further improved under the assumption that the length
and time scales of the newly created eddies are controlled by the
local-in-time friction velocity
\begin{equation}
u_\tau^{\star\, 2}(t)=\sqrt{ \left(\nu
  \frac{\partial \langle u_1 \rangle}{\partial x_2}\right)^2 + \left(\nu \frac{\partial \langle
  u_3 \rangle}{\partial x_2}\right)^2 }\bigg|_{x_2=0}. 
\end{equation}
The local wall units, denoted by $(\cdot)^\star$, are analogously
defined in terms of $\nu$ and $u_\tau^\star(t)$, and the local
friction Reynolds number is $Re_\tau^\star(t) = u_\tau^\star(t)
h/\nu$. The results in figure \ref{fig:scaling_inner}(c) confirm that
the local scaling ($t^\star$ versus $D_\tau Re^\star_\tau/\Pi$) holds
for longer times.
%

\begin{figure}
\vspace{0.5cm}
\begin{center}
   \vspace{0.1cm}
 \includegraphics[width=1\textwidth]{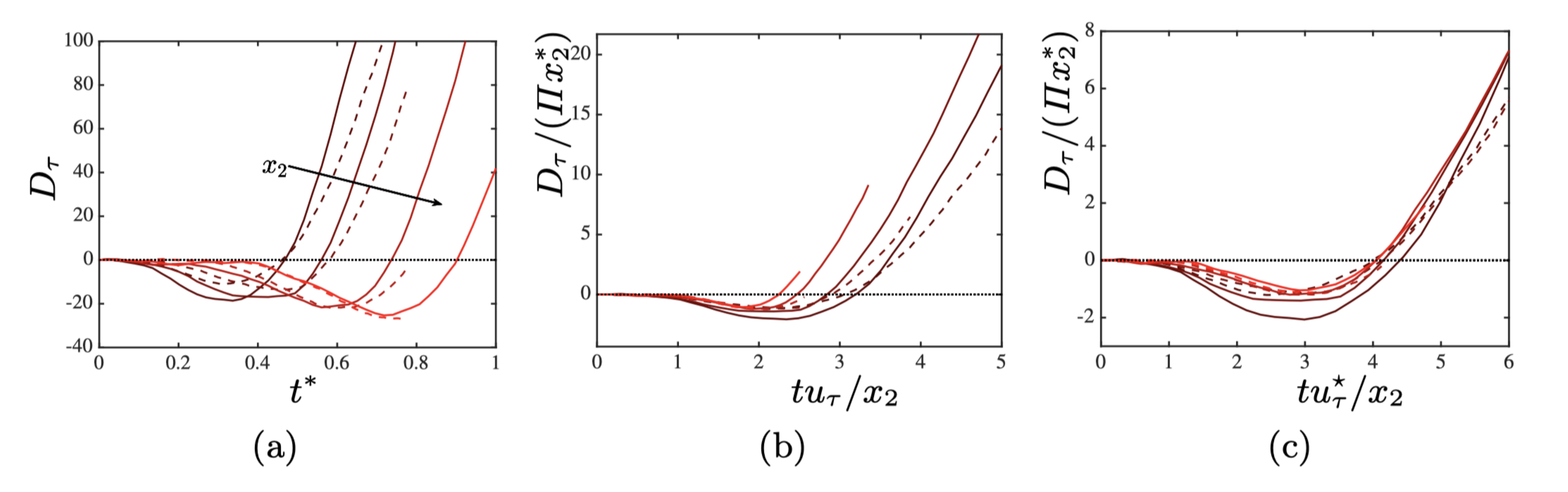}
 \end{center}
\caption{ Time evolution of the percentage change of tangential
  Reynolds stress $D_\tau$ in the logarithmic layer for
  $x_2^*=0.15,0.2,0.3$, and $0.4$ represented by lines coloured from
  dark red to light red.  Lines are (\solid) for cases at $Re_\tau
  \approx 500$ and (\dashed) for cases at $Re_\tau \approx 1000$, both
  for $\Pi =60$. The arrow in panel (a) indicates increasing
  wall-normal distance.
  \label{fig:scaling_log} }
\end{figure}

The time evolution of $D_\tau$ for the momentum-carrying eddies across
the logarithmic layer is shown in figure \ref{fig:scaling_log}, where
three scaling laws are investigated. The evolution of $D_\tau$ in
outer units is include in figure \ref{fig:scaling_log}(a).
Wall-attached eddies follow an ordered response in time after the
sudden imposition of the transverse pressure gradient: eddies closer
to the wall react earlier and are the least perturbed, while larger
eddies experience a more acute Reynolds stress reduction at later
times. The preceding analysis for the buffer region is extended to the
logarithmic layer by taking into consideration that the lifetimes of
the wall-attached eddies scale as $\sim$$x_2/u_\tau$, with a
consistent drop in the Reynolds stress proportional to $\Pi
x_2^*$. The self-similar response of wall-attached eddies under the
lateral force is evidenced by the improved collapse in figure
\ref{fig:scaling_inner}(b), at least for $t u_\tau/x_2 \lesssim
1$. Analogously to the inner layer, $u_\tau$ stands as the
characteristic velocity scale of the original eddies in the
equilibrium state, but does not hold as such for times longer than the
lifespan of individual wall-attached eddies, $t u_\tau/x_2 \approx 1$
\citep{Lozano2014b}.  The collapse among cases is perfected by using
the local time-scale $t u_\tau^\star/x_2$ (figure
\ref{fig:scaling_inner}c), which accounts for variations in the
momentum transfer controlling the eddies during the transient.
%

\subsection{Structural changes in the conditionally averaged flow field}
\label{subsec:structure}

\begin{figure}
\begin{center}  
   \vspace{0.5cm}
  \includegraphics[width=1\textwidth]{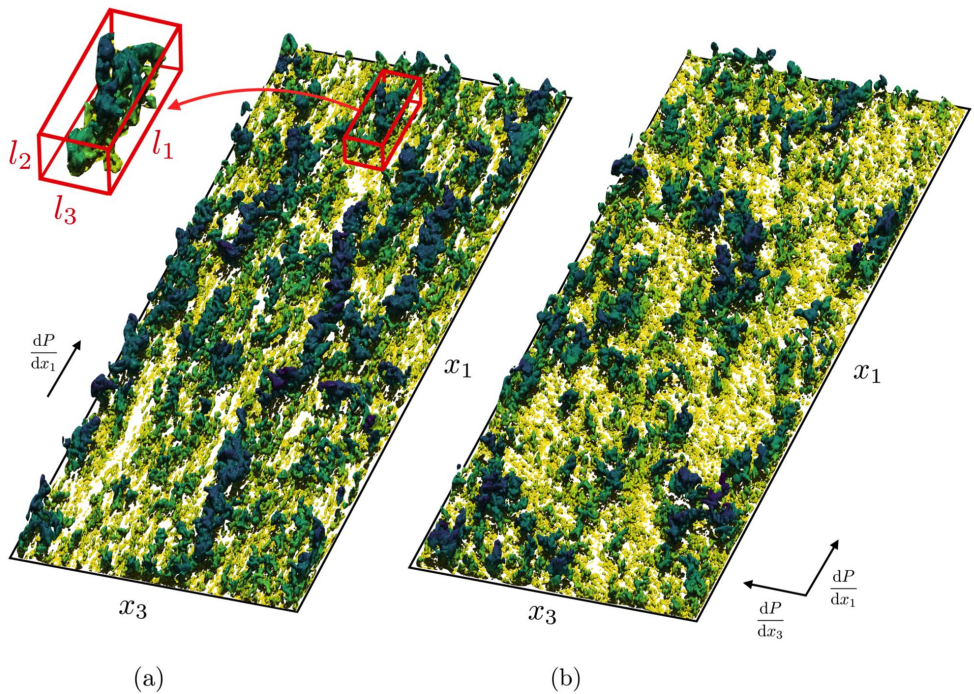}
 \end{center}
\caption{ Instantaneous $-u_1' u_2'$-structures defined by
  (\ref{eq:thr:uvs}) for $\Pi =60$ and $Re_\tau\approx 1000$ at (a)
  $t^*=0$ and (b) $t^*=0.5$. Only $-u_1' u_2'$-structures attached to
  the bottom wall are shown. The colours represent the distance to the
  wall from yellow (closer to the wall) to blue (farther from the
  wall). The box with edges coloured in red is the bounding box of one
  individual $-u_1' u_2'$-structure with streamwise, wall-normal, and
  spanwise sizes equal to $l_1$, $l_2$, and $l_3$,
  respectively. \label{fig:uv_snapshot} }
\end{figure}

We examine the structural evolution of the flow in the surroundings of
the momentum-carrying eddies. To that end, we identify
three-dimensional structures of the intense momentum transfer using
the methodology introduced by \cite{Lozano2012} \citep[see
  also][]{Lozano2014b,Lozano2016}.  An individual structure (or
object) of intense momentum transfer at time $t$ is defined as a
spatially connected region in the flow satisfying
\begin{equation}\label{eq:thr:uvs}
-u_1'(x_1,x_2,x_3,t) u_2'(x_1,x_2,x_3,t) > 
H \langle u'^2_1 \rangle^{1/2}(x_2,t) \langle u'^2_2\rangle^{1/2}(x_2,t),
\end{equation}
where $H$ is a thresholding parameter \citep[hyperbolic-hole
  size,][]{Bogard1986} equal to $1.75$ obtained following the analysis
by \cite{Moisy2004}. It was tested that varying $H$ within the range
$0.5<H<3$ does not change the conclusions below. The original frame of
reference defined by $\mathcal{F}$ is preferred to
$\tilde{\mathcal{F}}$ in order to avoid artificial distortions in the
flow due to the time and space variations in $\tilde{\mathcal{F}}$.
Hereafter, we refer to individual structures of intense $-u_1' u_2'$
events as $-u_1' u_2'$-structures. Numerically, three-dimensional
structures are constructed by connecting neighbouring grid points
fulfilling (\ref{eq:thr:uvs}) and using the 6-connectivity criteria
\citep{Rosenfeld1982}. Figure \ref{fig:uv_snapshot} shows the
wall-attached $-u_1' u_2'$-structures identified before and after the
imposition of the spanwise pressure gradient.  Figure
\ref{fig:uv_snapshot} also includes one individual $-u_1'
u_2'$-structure highlighted by the box with red edges.
%

\begin{figure}
\begin{center}  
   \vspace{0.1cm}
 \includegraphics[width=1\textwidth]{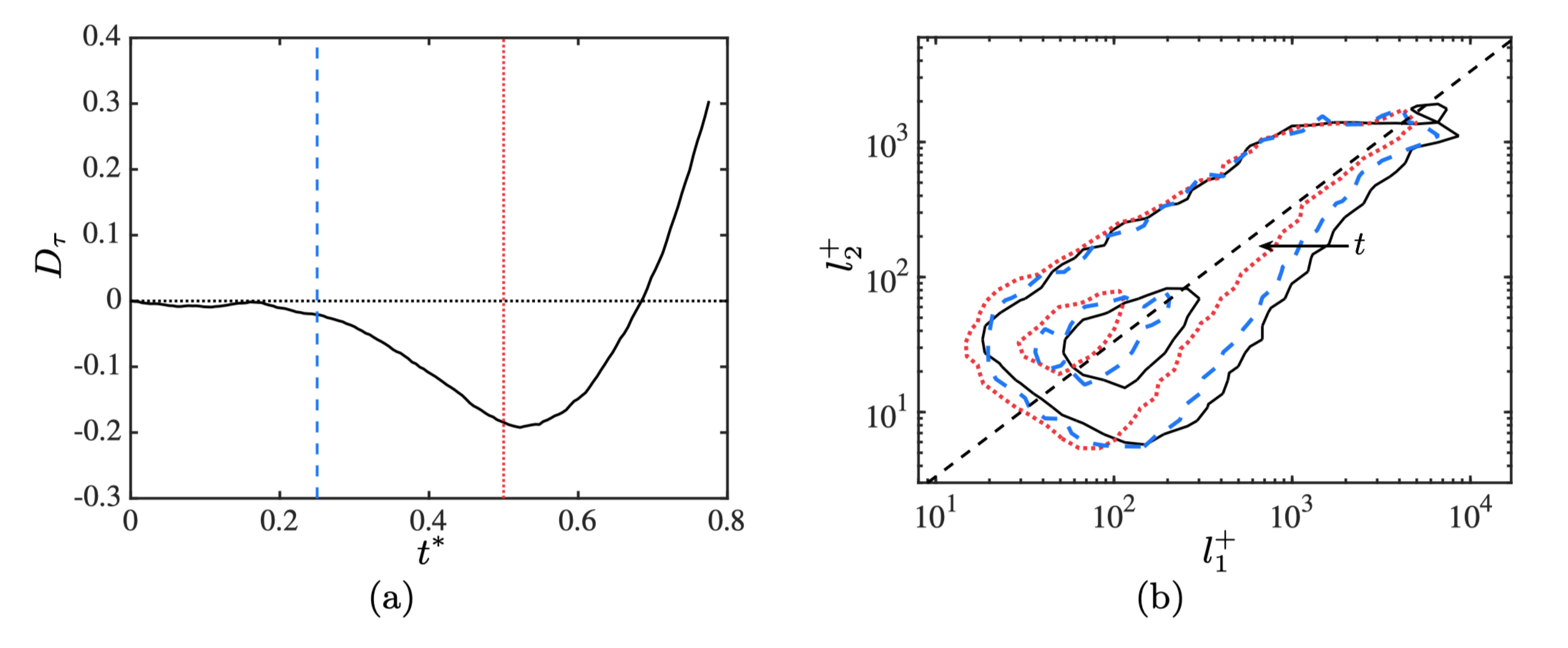}
 \end{center}
\caption{ (a) Time evolution of the percentage change of tangential
  Reynolds stress $D_\tau$ at $x_2^*=0.25$. The vertical lines are the
  times selected to study the flow structure in addition to $t^*=0$,
  namely $t^*=0.25$ (\dashed) and $t^*=0.5$ (\dotted). (b) Joint
  probability density functions of the logarithms of the streamwise
  $l_1$ and wall-normal $l_2$ sizes of wall-attached $-u_1'
  u_2'$-structures, $p(l_1^+,l_2^+)$.  The contours plotted contain
  50\% and 99.8\% of the probability. The lines are
  (\textcolor{black}{\solid}) for $t^*=0$, (\textcolor{blue}{\dashed})
  for $t^*=0.25$, and (\textcolor{red}{\dotted}) for $t^*=0.5$. The
  straight dashed line is $l_1^+ = 3l_2^+$ and the arrow indicates the
  direction of time. The results are for $Re_\tau\approx 1000$ and
  $\Pi=60$.
\label{fig:time_uv_size} }
\end{figure}

We focus our attention on the channel at $Re_\tau\approx 1000$ and
$\Pi=60$, but similar results are obtained consistently across
different $Re_\tau$ and $\Pi$ provided that the latter is large
enough to attain the fully non-equilibrium regime. We select three
time instants to assess the structural changes in the flow, namely,
$t^*=0$, $t^*=0.25$, and $t^*=0.50$. The time evolution of $D_\tau$ is
plotted in figure \ref{fig:time_uv_size}(a), which shows that the
maximum drop in the tangential Reynolds stress occurs at $t^*\approx
0.50$.
%

The identification procedure above yields about $10^5$ structures at
each time instant after discarding those objects with volumes smaller
than $30^3$ wall units.  The sizes of the objects are measured by
circumscribing each structure within a box aligned to the Cartesian
axes, whose streamwise, wall-normal, and spanwise sizes are denoted by
$l_1$, $l_2$, and $l_3$, respectively.  The minimum and maximum
distances of each object to the closest wall are $x_{2,\min}$ and
$x_{2,\max}$, respectively, and such that $l_2 =
x_{2,\max}-x_{2,\min}$.  An example of an individual $-u_1'
u_2'$-structure and its bounding box is included in figure
\ref{fig:uv_snapshot}(a). We centre our attention on wall-attached
$-u_1' u_2'$-structures, defined as those with $x_{2,\min}^+<25$
\citep{DelAlamo2006b}. For the value of $H$ selected, wall-attached
structures are responsible for more than 60\% of the tangential
Reynolds stress at all three times considered. Figure
\ref{fig:time_uv_size}(b) shows the joint probability density function
(p.d.f.) of the sizes of the wall-attached structures,
$p(l_1^+,l_2^+)$. At $t^*=0$, the distribution of sizes is consistent
with a geometrically self-similar population of structures akin to the
wall-attached eddies envisioned by \citet{Townsend1976}. The mode of
the p.d.f. follows a reasonably well-defined linear law, $l_1 \sim 3
l_2$ consistent with previous studies \citep{Lozano2012}. From $t^*=0$
to $t^*=0.50$, the most pronounced modification in the geometry of the
structures is a gradual shortening of their streamwise length, while
their wall-normal heights are barely affected.

Each $-u'_1u'_2$-structure can be classified as either an ejection,
when the average wall-normal velocity within its enclosed volume is
positive, or as a sweep otherwise.  Sweeps and ejections are known to
be spatially organised in pairs side-by-side along the spanwise
direction \citep{Ganapathisubramani2008, Lozano2012, Wallace2016,
  Kosuke2018}. This sweep-ejection group, representative of a
streamwise roll, is the predominant logarithmic-layer flow structure
responsible for the generation of tangential Reynolds
stress. Consequently, we are interested in examining the modification
of the flow around sweep-ejection pairs during the transient
period. We denote the centre of gravity of the bounding boxes of the
$n$-th sweep and its paring ejection as $\xvec_s^n$ and $\xvec_e^n$,
respectively. The wall-normal size of the sweep is $l_{2,s}^n$ and of
the ejection $l_{2,e}^n$.  The averaged flow field conditioned to the
presence of a sweep-ejection pair is computed by averaging the
velocity vector in a rectangular domain along different $n$-th pairs,
whose centre coincides with $\xvec_p^n=(\xvec_e^n+\xvec_s^n)/2$, and
it edges are $\boldsymbol{r}$ times the average wall-normal height
$l_p^n = (l_{2,e}^n+l_{2,s}^n)/2$. Then, the conditionally averaged
flow around sweep-ejection pairs is given by
\begin{equation}
\{u'_i\}(\boldsymbol{r})  = 
\sum_{n=1}^{N} \frac{u'_i(\xvec_p^n + l_p^n \boldsymbol{r})}{N},
\la{eq:condave}
\end{equation}
where $n=1,..,N$ is the set of sweep-ejection pairs selected to
perform the conditional average, and
$\boldsymbol{r}=(r_1,r_2,r_3)$. We also take advantage of the spanwise
symmetry of the flow, and $r_3$ is always chosen to be positive
towards the sweep.  The reader is referred to \cite{Lozano2012} and
\cite{Dong2017} for additional details on the procedure to obtain
conditional flow fields.

\begin{figure}
\vspace{0.5cm}
\begin{center} 
 \includegraphics[width=1\textwidth]{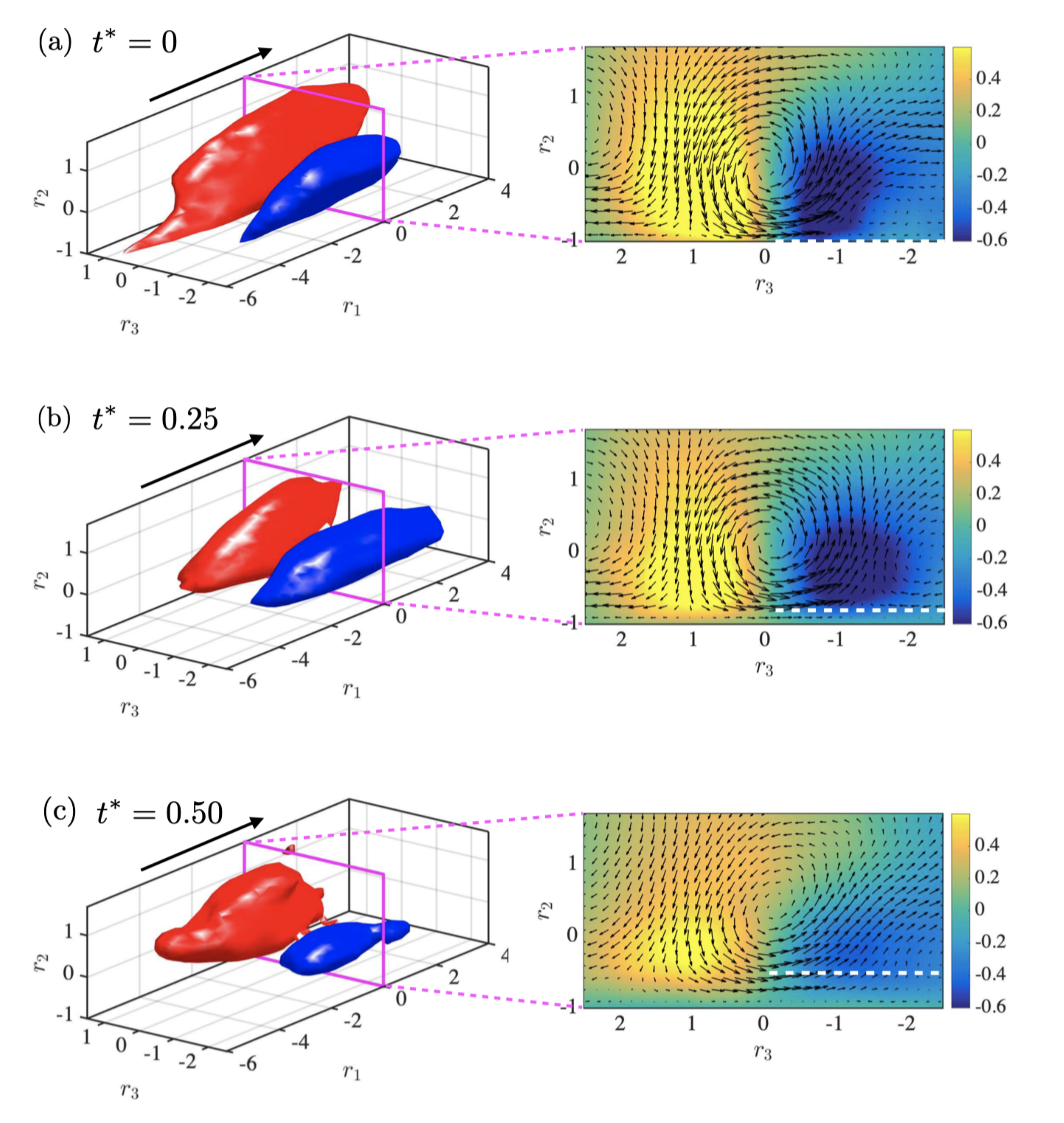}
 \end{center}
\caption{ Averaged flow fields conditioned to wall-attached pairs of
  sweeps and ejections with wall-normal sizes in the range
  $0.2<l_p^{n\,*}<0.3$ at (a) $t^*=0$, (b) $t^*=0.25$, and (c)
  $t^*=0.50$. Panels on the left contain isosurfaces of the low-
  (blue) and high- (red) velocity streaks defined by $\pm \alpha$ of
  the maximum positive and negative, respectively, fluctuating
  streamwise velocity of the average flow with (a) $\alpha=0.6$, (b)
  $\alpha=0.55$, and (c) $\alpha=0.43$. The arrows indicate the mean
  flow direction. Panels on the right display the cross-flow velocity
  vector field $(\{u_2'\},\{u_3'\})$ (arrows) and the fluctuating
  velocity $\{u_1'\}$ (colours). The dashed white line shows the
  wall-normal extension from the wall of the incoherent streamwise
  velocity field represented by low values of $\{u_1'\}$ in green.
  Velocities are normalised by $u_\tau$. Results are for $\Pi=60$ at
  $Re_\tau \approx 1000$.  \label{fig:flow_structure} }
\end{figure}

The averaged flow field conditioned to sweep-ejection pairs with $0.2<
l^{n\,*}_p< 0.3$ is plotted in figure \ref{fig:flow_structure}. At
$t^*=0$ (figure \ref{fig:flow_structure}a), the characteristic flow
structure consistent with the statistically in equilibrium state is a
streamwise roll flanked by one low-velocity streak and one
high-velocity streak.  At succeeding times (figure
\ref{fig:flow_structure}b,c), the roll persist, while the intensity
and size of the low-velocity streak decrease. The high-velocity
streaks and roll are also weakened, but the variations are less
pronounced. The second observation is the loss of coherence in a
developing layer underneath the low-velocity (green region of low
streamwise fluctuating velocity below the white dashed lines in figure
\ref{fig:flow_structure}). During the transient, both low- and
high-velocity streaks shorten in the streamwise direction in
accordance with the geometric analysis in figure
\ref{fig:time_uv_size}(b). Although not shown, the results above are
also applicable to sweep-ejection pairs across different ranges of
$l_p^n$ when the times are appropriately scaled by
$l_p^n/u^\star_\tau$, i.e. the modifications in the flow are
self-similar in space and time.
%

\begin{figure}
\vspace{0.5cm}
\begin{center}  
 \includegraphics[width=1\textwidth]{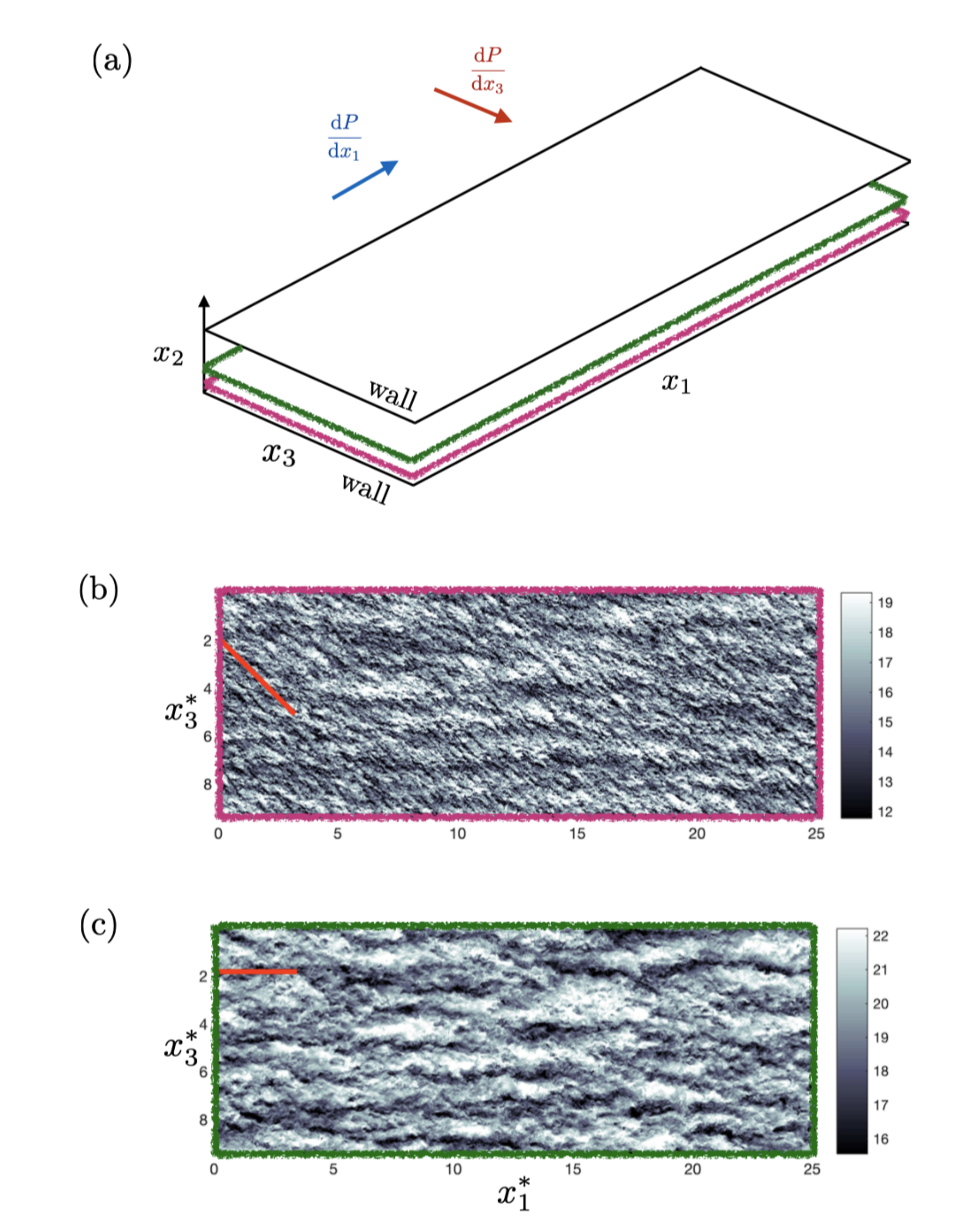}
 \end{center}
\caption{ (a) Schematic of the channel flow domain with the
  wall-normal planes shown in panels (b) and (c). Panels (b) and (c)
  contain the streamwise velocity at (b) $x_2^*=0.1$ and (c)
  $x_2^*=0.3$ at the same time instant $t^*=0.5$ for $\Pi=60$ at
  $Re_\tau \approx 1000$. The red straight lines in (b) and (c)
  indicate the mean flow direction.  Velocities are normalised by
  $u_\tau$.\label{fig:snapshots_3D} }
\end{figure}

The message from figure \ref{fig:snapshots_3D} is that the main
structural alteration during the transient is the weakening of the
low-velocity streaks, which is in turn associated with the loss of
coherence of the flow within a growing layer underneath the streamwise
rolls. The aforementioned loss of coherence may be the consequence of
the relative displacement of wall-parallel layers at different heights
and the additional mean spanwise shear which enhances the generation
of smaller flow scales. This is illustrated in figure
\ref{fig:snapshots_3D}, which contains the instantaneous streamwise
velocity at two wall-normal distances; one closer to the wall at
$x_2^*=0.1$ influenced by the additional shear from the lateral
boundary layer, and another farther from the wall at $x_2^*=0.3$ still
unaffected.
%

\subsection{Structural model}

\begin{figure}
\begin{center}  
   \vspace{0.3cm}
 \includegraphics[width=0.97\textwidth]{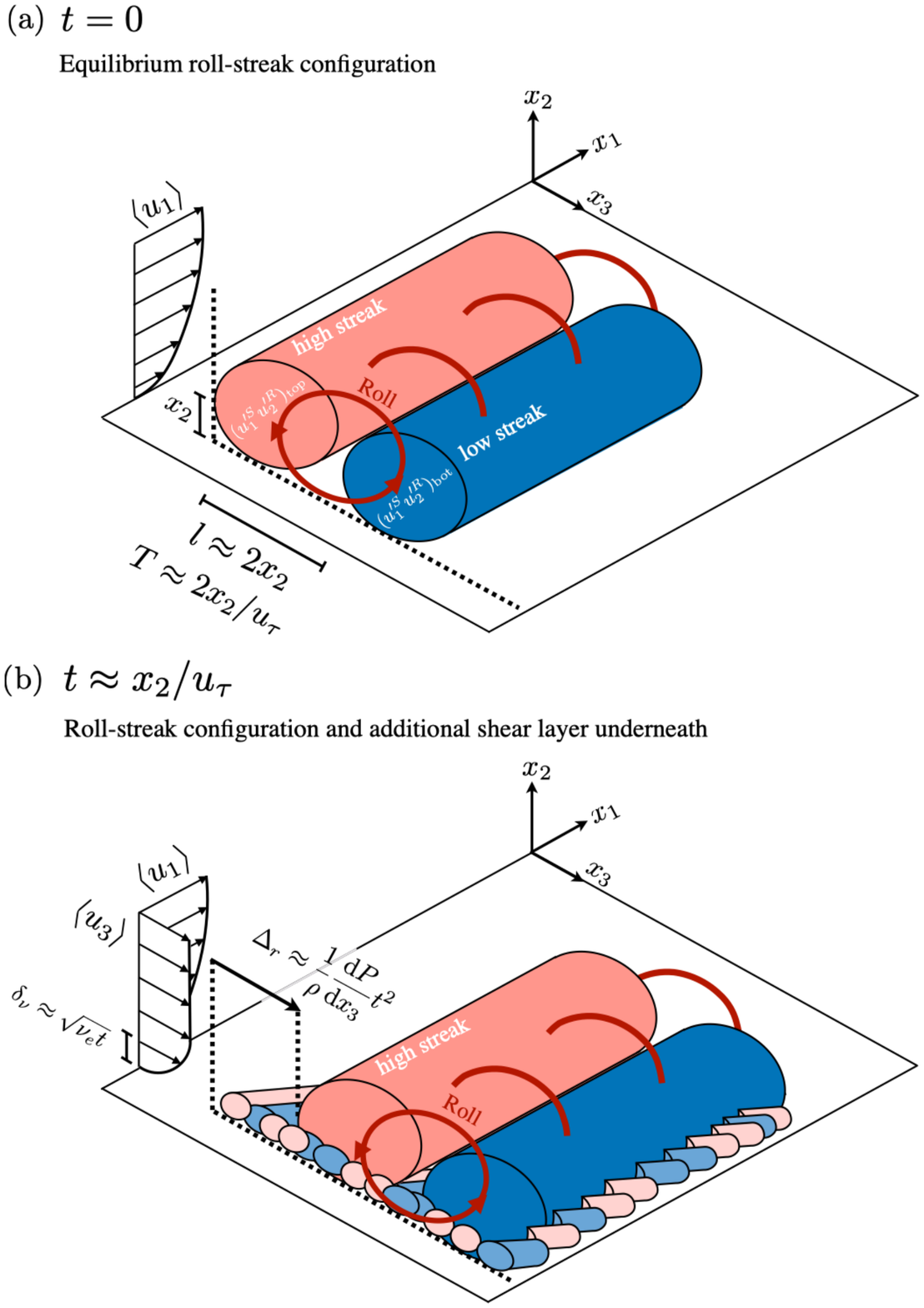}
 \end{center}
\caption{ Structural model of self-similar wall-attached eddies
  subjected to a sudden mean spanwise pressure gradient. The figure
  shows one building block structure which comprises a streamwise roll
  flanked by one low-velocity streak and one high-velocity streak. (a)
  Statistically in equilibrium wall-attached momentum-carrying eddies
  of size $2x_2$ at $t=0$ generating a momentum transfer
  $\approx$$-u_\tau^2$. (b) Non-equilibrium wall-attached
  momentum-carrying eddies at $t=x_2/u_\tau$ after the imposition of a
  transverse mean pressure gradient generating a momentum transfer
  $\approx$$-u_\tau^2/2(1+\lambda)$.  In panel (a), $(u'^S_1
  u'^R_2)_{\mathrm{top}}$ and $(u'^S_1 u'^R_2)_{\mathrm{bot}}$
  represent the downward and upward, respectively, wall-normal
  momentum transfer by the streamwise roll. In panel (b), $\delta_\nu$
  is the spanwise boundary layer thickness and $\Delta_r$ is the
  lateral displacement of the flow above $\delta_\nu$ due to the
  uniform acceleration $(1/\rho)\mathrm{d}P/\mathrm{d}x_3$ imposed by
  the mean spanwise pressure gradient.
\label{fig:model_sketch.eps} }
\end{figure}

On the basis of the above observations, we propose a conceptual model
that accounts for the changes undergone by the flow. The model is
sketched in figure \ref{fig:model_sketch.eps}. The key elements are
the low- and high-velocity streaks and their relative alignment with
respect to the streamwise roll. At a given wall-normal distance $x_2$
and $t=0$, the flow is configured in an equilibrium array of rolls and
streaks with their centres at $x_2$, sizes $2x_2$, and lifetimes $2
x_2/u_\tau$ \citep{Lozano2014b}. The tangential Reynolds stress
$\langle u'_1 u'_2 \rangle$ at $x_2$ is the result of the wall-normal
momentum transport conducted by the rolls and the arrangement of
streaks in the equilibrium state. The momentum transfer at $t=0$ can
be modelled as the sum of two contributions,
\begin{eqnarray}\label{eq:model:uv}
\langle u'_1 u'_2 \rangle(x_2,t=0)^{\mathrm{model}}
&\approx&  (u'^S_1 u'^R_2)_{\mathrm{top}} + (u'^S_1 u'^R_2)_{\mathrm{bot}} \\
&\approx& (u_\tau) (-u_\tau/2) + (-u_\tau) (u_\tau/2) \approx -u_\tau^2,
\end{eqnarray}
where $(u'^S_1 u'^R_2)_{\mathrm{top}}$ represents the wall-normal
transport of the high-velocity streak, $u'^S_1\approx u_\tau$, by
the downward motion of the roll, $u'^R_2 \approx -u_\tau/2$, above
$x_2$. Conversely, $(u'^S_1 u'^R_2)_{\mathrm{bot}}$ is the wall-normal
transport of the low-velocity streak, $u'^R_1\approx -u_\tau/2$, by
the upward motion of the roll, $u'^R_2 \approx u_\tau$, below
$x_2$. The intensities of $u'^S_1$ and $u'^R_2$ are adjusted to
produce a total momentum transfer equal to $-u_\tau^2$, although the
discussion is extensive to other values.
%

At half the lifespan of the eddies $t\approx x_2/u_\tau$, the spanwise
boundary layer extends up to $\delta_\nu \approx 0.445 x_2$, based on
the estimations in \S\ref{sec:overview}, and remains below the centre
of the rolls located at $x_2$. Simultaneously, the upper flow is
laterally displaced by $\Delta_r \approx
(1/\rho)\mathrm{d}P/\mathrm{d}x_3 (x_2/u_\tau)^2$.  For values of
$\Delta_r$ larger than the spanwise coherence of the roll-streak
structure, namely $\Delta_r > 2 x_2$, the centre of the rolls is
misaligned with the underneath streaks within the lateral boundary
layer. The latter streaks are also altered by $\partial\langle
u_3\rangle /\partial x_2$, which increases the local Reynolds number
and triggers the emergence of smaller scales. These changes originate
a new flow configuration which is less efficient in producing $\langle
u'_1 u'_2 \rangle$ compared to the equilibrium state. The new momentum
transfer at $t\approx x_2/u_\tau$ can be modelled similarly to
(\ref{eq:model:uv}) by assuming that $(u'^S_1 u'^R_2)_{\mathrm{top}}$
is barely affected, whereas $(u'^S_1 u'^R_2)_{\mathrm{bot}}$ provides
a deficient momentum transfer such that
\begin{eqnarray}\label{eq:model:uv_non}
\langle u'_1 u'_2 \rangle(x_2,t=x_2/u_\tau)^{\mathrm{model}} &\approx& 
(u'^S_1 u'^R_2)_{\mathrm{top}} + (u'^S_1 u'^R_2)_{\mathrm{bot}} \\
&\approx&  (u_\tau) (-u_\tau/2) + (-u_\tau \lambda) (u_\tau/2) \approx -\frac{u_\tau^2}{2}(1 + \lambda),
\end{eqnarray}
where $\lambda$ is a damping factor accounting for the reduction in
the Reynolds stress generation due to the loss of streak coherence
within the lateral boundary layer. The functional form of $\lambda$ is
modelled by assuming that the loss of streak coherence is, in first
order approximation, linearly proportional to the relative spanwise
mean shear,
\begin{equation}\label{eq:f}
\lambda = 1 -
\frac{\partial\langle u_3\rangle /\partial x_2}{\partial\langle u_1\rangle /\partial x_2},
\end{equation}
such that $\langle u'_1 u'_2 \rangle= -u_\tau^2$ for $\partial\langle
u_3\rangle /\partial x_2 = 0$. If we consider the approximations
$\partial\langle u_3\rangle /\partial x_2 \approx
\mathrm{d}P/\mathrm{d}x_3 / (\rho u_\tau)$ and $\partial\langle
u_1\rangle /\partial x_2 \approx u_\tau/(2 x_2)$ (see
\S\ref{sec:overview}), then
\begin{equation}\label{eq:model:uv_final}
\langle u'_1 u'_2 \rangle(x_2,t=x_2/u_\tau)^{\mathrm{model}} 
\approx -u_\tau^2\left( 1 - \frac{x_2\mathrm{d}P/\mathrm{d}x_3}{\rho u_\tau^2}\right).
\end{equation}
Equation (\ref{eq:model:uv_final}) can be re-arranged as
$\min_t\{D_\tau^{\mathrm{model}}\} \approx -\Pi x_2^*$, which
coincides with the maximum Reynolds stress depletion from
(\ref{Dtau_log}).

Additionally, the model above predicts that the condition for
non-equilibrium of flow structures at height $x_2$ is given by
$\Delta_r > 2 x_2$, which in non-dimensional form yields $\Pi
x_2^*>2$. In order to disturb the wall-parallel layers at all heights
across the boundary layer, $x_2^*$ should be fixed in wall units and
such that $\Pi > \mathcal{O}(Re_\tau)$, also consistent with the
estimation of $\Pi > 0.03Re_\tau$ provided in \S\ref{subsec:regimes}.

The scenario promoted above is self-similar: the continuous depletion
in time of the Reynolds stress in figure \ref{fig:stress_aligned}(b)
is the result of the time-ordered disruption of streaks and rolls from
their natural equilibrium by the growth of the spanwise boundary
layer. The above mechanism also shares some similarities with the
physical arguments pertaining to the modification of near-wall
turbulence in the presence of oscillating walls characteristic of drag
reduction studies \citep{Jung1992, Laadhari1994, Choi2001, Choi2002,
  Ricco2008}, although our model is tailored for multiscale flows
and uniform accelerations.

Finally, it is worth noting that the reduction of the Reynolds stress
has been mainly modelled on the basis of non-equilibrium effects
rather than on the three-dimensionality of the mean flow and,
therefore, is not constrained to the application of additional mean
pressure gradients only in the spanwise direction. Accordingly, the
model also predicts that a sudden forcing in the streamwise direction
would encompass a decrease of the Reynolds stress as long as the
relative shift between wall-parallel layers is capable of misaligning
the cores of the roll and the streaks underneath. As it is well-known
that streaks are longer than wider across the logarithmic layer by a
factor of 3 to 6, we can anticipate that suddenly imposed streamwise
mean pressure gradients are less efficient in decreasing the Reynolds
stress than their spanwise counterparts. This view is supported by the
studies by \cite{He2013}, \cite{He2015} \cite{Seddighi2015}, and
\cite{Mathur2018}, who showed that channel flows subjected to
streamwise mean pressure gradients exhibit a similar, but less
exacerbated, counter-intuitive response of flow consistent with the
model presented here.

\subsection{Time evolution of the tangential Reynolds stress budget}
\label{subsec:budget}

We examine the reduction of $-\langle u_1' u_2' \rangle$ from the
Reynolds-stress budget viewpoint to complement the physical insight
gained from the structural analysis in \S\ref{subsec:structure}.  We
use the static frame of reference $\mathcal{F}$ to avoid the
complexity of additional terms of the form $\partial/\partial t$ in
the budget equation. Following \cite{Mansour1988}, the dynamic
equation for the component $\langle u_i' u_j'\rangle$ is given by
\begin{equation}\label{eq:budget}
\frac{\mathcal{D} \langle u_i' u_j'\rangle}{\mathcal{D}t} = P_{ij}
+ \varepsilon_{ij} + T_{ij} + PS_{ij} + PD_{ij} + V_{ij},
\end{equation}
where the terms in the right-hand side of (\ref{eq:budget}) are the
Reynolds stress production ($P_{ij}$), dissipation
($\varepsilon_{ij}$), turbulent diffusion ($T_{ij}$), pressure strain
($PS_{ij}$), pressure diffusion ($PD_{ij}$), and viscous diffusion
($V_{ij}$) defined as
\begin{eqnarray}\label{eq:budget_terms}
P_{ij}   &=& 
-\langle u_i' u_k' \rangle \left\langle \frac{\partial u_j}{\partial x_k}\right\rangle
-\langle u_j' u_k' \rangle \left\langle \frac{\partial u_i}{\partial x_k}\right\rangle, \\
\varepsilon_{ij}  &=& -2\nu \left\langle \frac{\partial u_i'}{\partial x_k}  \frac{\partial u_j'}{\partial x_k} 
\right\rangle, \\
T_{ij}   &=& -\left\langle \frac{\partial u_i' u_j' u_k'}{\partial x_k} \right\rangle, \\
PS_{ij}  &=& -\left\langle u_i' \frac{\partial p'}{\partial x_j} + u_j' \frac{\partial p'}{\partial x_i} 
\right\rangle, \\
V_{ij}   &=&  \left\langle \frac{\partial^2 u_i' u_j'}{\partial x_k \partial x_k} \right\rangle. 
\end{eqnarray}
In order to obtain quantities that are only a function of time, we
introduce the average along $x_2$-bands, which is indicated by
$\overline{(\cdot)}$. The wall-normal bands inspected are $x_2^+ \in
[5,50]$ and $x_2^* \in [0.2,0.3]$, which lie within the buffer region
and logarithmic layer, respectively.  The gains produced by the budget
components $\bar\phi_{ij}$ for $(i,j)=(1,2)$, $(i,j)=(1,1)$, and
$(i,j)=(2,2)$ are defined as
\begin{eqnarray}
\mathrm{Gain}_{-12} &=& \frac{-\bar\phi_{12}(t) + \bar\phi_{12}(0)}{-\overline{P}_{12}(0)}, \\
\mathrm{Gain}_{11} &=& \frac{\bar\phi_{11}(t) - \bar\phi_{11}(0)}{\overline{P}_{11}(0)}, \\
\mathrm{Gain}_{22} &=& \frac{\bar\phi_{22}(t) - \bar\phi_{22}(0)}{\overline{PS}_{22}(0)}, 
\end{eqnarray}
where $\mathrm{Gain}_{-12}$, $\mathrm{Gain}_{11}$, and
$\mathrm{Gain}_{22}$ represent the gain in the Reynolds-stress budget
equation for $-\overline{\langle u_1'u_2' \rangle}$,
$\overline{\langle u_1'u_1'\rangle}$, and $\overline{\langle
  u_2'u_2'\rangle}$, respectively.  Note that $\mathrm{Gain}_{-12}$
is defined such that $-\overline{P}_{12}>0$ contributes to increasing
the magnitude of $-\overline{\langle u_1'u_2'\rangle}$. We analyse
the channel flow at $Re_\tau \approx 500$ and $\Pi = 60$, in which
case the maximum drop in $-\overline{\langle u_1'u_2' \rangle}$
occurs at $t^+\approx 170$ and $t^*\approx0.7$ for the bands in the
buffer region and logarithmic layer, respectively.
%
\begin{figure}
\vspace{0.5cm}
\begin{center}
%
 \includegraphics[width=1\textwidth]{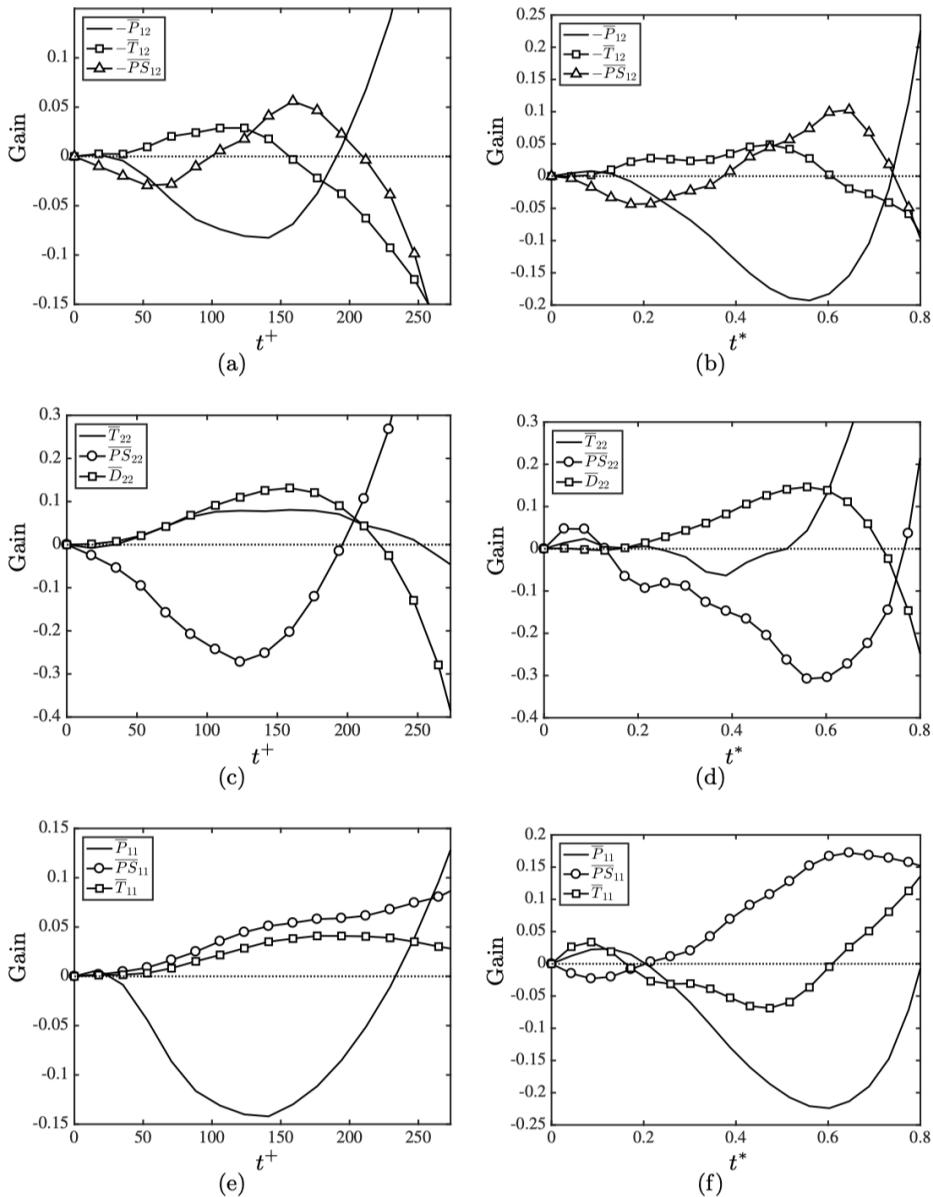}
 \end{center}
\caption{ Time evolution of the gain produced by the different terms
  of the Reynolds-stress budget for (a,b) $-\overline{\langle
    u_1'u_2'\rangle}$, (c,d) $\overline{\langle u_2'u_2'\rangle}$, and
  (e,f) $\overline{\langle u_1'u_1'\rangle}$. Panels on the left are
  for $x_2^+ \in [5,50]$, and panels on the right are for $x_2^* \in
  [0.2,0.3]$. The results are for $\Pi = 60$ at $Re_\tau \approx
  500$. Zero gain is represented by (\dotted).
  \label{fig:budget} }
\end{figure}

The gains are reported in figure \ref{fig:budget} as a function of
time.  We discuss first the results for the buffer layer region
$x_2^+ \in [5,50]$.  Figure \ref{fig:budget}(a) shows the
time evolution of $-\overline{P}_{12}$, $-\overline{T}_{12}$, and
$-\overline{PS}_{12}$. The remaining terms are not significant in
magnitude nor they play any relevant role in the discussion below and
they are omitted for the sake of simplicity. The main contributor to
the destruction of $-\overline{\langle u_1' u_2'\rangle}$ is the drop
in production $-\overline{P}_{12}$, which can be traced back to a
deficit on the pressure-strain correlation in the budget equation for
$\overline{\langle u_2' u_2'\rangle}$ (figure
\ref{fig:budget}c). Similarly, the decline of the streaks is the
consequence of a lower production $\overline{P}_{11}$ (figure
\ref{fig:budget}e), also caused by the drop in $-\overline{\langle
  u_1' u_2'\rangle}$.

The sequence of events is similar farther away from the wall as seen
in figures \ref{fig:budget}(b,d,f) for $x_2^* \in [0.2,0.3]$.  The
main sink of tangential Reynolds stress arises from the turbulent
production $-\overline{P}_{12}$.  The reduction in
$-\overline{P}_{12}$ is connected to the lower pressure-strain
correlation $\overline{PS}_{22}$ in the budget equation of
$\overline{\langle u'_2u'_2\rangle}$ akin to the situation described
for the buffer region.  The decay of the streaks is similarly governed
by the drop in the production of streamwise Reynolds stress
$\overline{P}_{11}$, with some additional contribution by the
turbulent diffusion $\overline{T}_{11}$.

\begin{figure}
\begin{center}  
   \vspace{0.6cm}
   \includegraphics[width=0.95\textwidth]{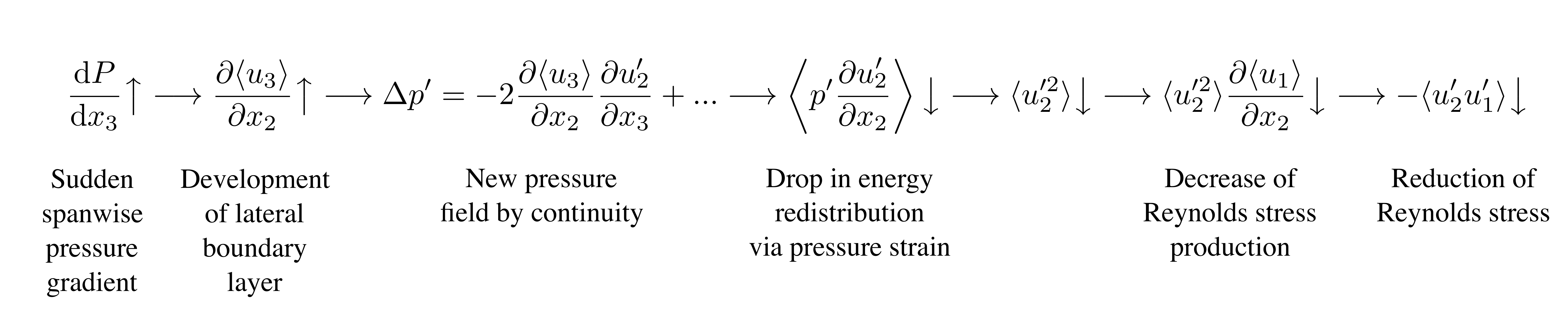} 
 \end{center}
\caption{Summary of the sequential process of Reynolds stress
  reduction from the sudden imposition of a spanwise pressure gradient
  up to the final decrease in tangential Reynolds stress. Time goes
  from left to right. See text for details.
\label{fig:steps} }
\end{figure}

The process of Reynolds stress reduction is then sequentially
described by \emph{(i)} the growth of the spanwise boundary layer
$\partial \langle u_3 \rangle/\partial x_2$, which \emph{(ii)}
inhibits the redistribution of energy to $\langle u'_2 u_2' \rangle$
via pressure-strain correlation, followed by \emph{(iii)} the
weakening of the production of tangential Reynolds stress, which
\emph{(iv)} eventually causes the drop in $-\langle u'_1 u_2'
\rangle$.  The terms involved at each step of the process are
summarised in figure \ref{fig:steps}. A similar effect has been
observed in transitional boundary-layer flows subjected to spanwise
wall oscillations \citep{Hack2015}.  Our findings are consistent with
the previous theory on transversely strained boundary-layer flows by
\cite{Moin1990} and \cite{Coleman1996a} and extends the results to the
outer layer of wall bounded turbulence. The leading role of $\partial
\langle u_3 \rangle /\partial x_2$ in the drop of $-\langle u_1'
u_2'\rangle$ is also consistent with the structural model promoted in
\S\ref{subsec:structure}, where it was argued that the deficient
transport of momentum by the streamwise rolls has its origin on the
displacement among fluid layers induced by $\partial \langle u_3
\rangle /\partial x_2$.
%

\section{Applications to wall-modelled LES}\label{sec:wmles}

\begin{figure}
\begin{center}
\vspace{0.5cm}
 \includegraphics[width=1\textwidth]{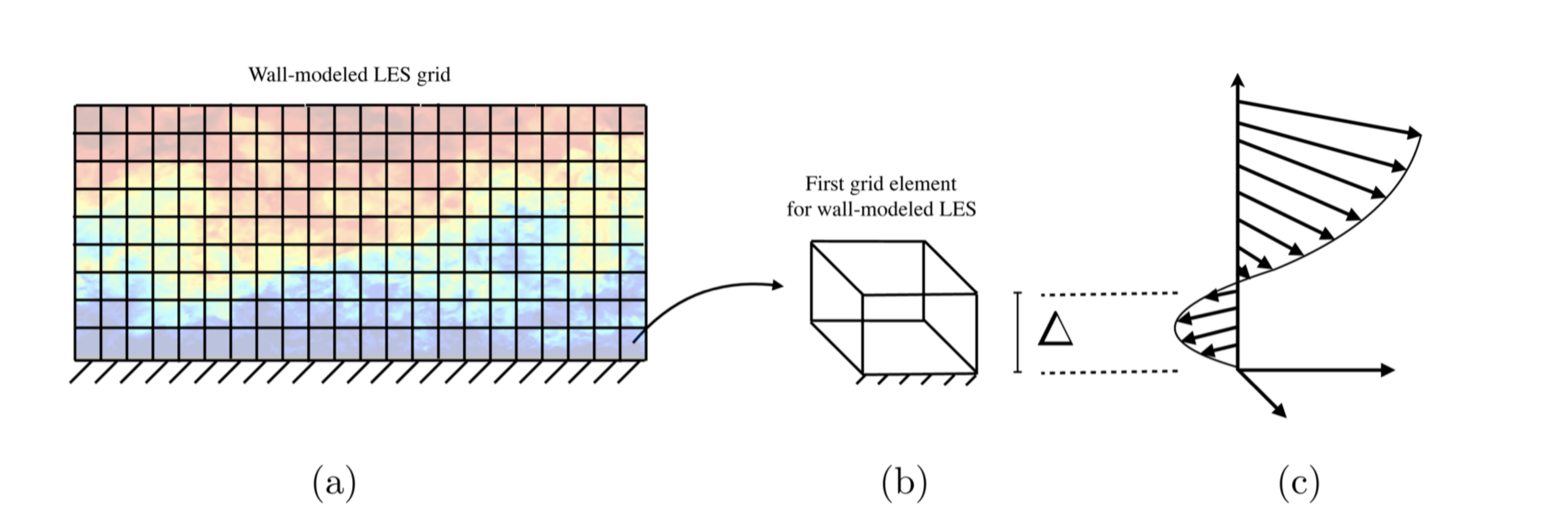}
\end{center}
\caption{(a) Sketch of a typical grid for WMLES of a turbulent
  boundary layer. The background colours represent the streamwise
  velocity from zero velocity (dark blue) to the free stream value
  (dark red). (b) First grid cell of size $\Delta$ at the wall, and
  (c) comparison with the potential directional changes in the mean
  velocity profile along the wall-normal coordinate in 3DTBL.
\label{fig:wmles_sketch}}
\end{figure}

We study the predictive capabilities of WMLES in non-equilibrium 3-D
channel flows at $Re_\tau\approx 1000$.  As discussed in previous
sections, this relatively simple flow set-up entails fundamental
features of 3DTBL that may challenge the available model
formulations. The rapid temporal and wall-normal variations in the
strain and vorticity, as illustrated in figure \ref{fig:wmles_sketch},
have the potential of rendering turbulence closure models calibrated
to equilibrium turbulence of limited utility.  Additionally, the
accurate prediction of the wall-shear angle and Reynolds stress
magnitude is also of paramount importance in external flows over wings
or bluff bodies, as it can directly affect the force exerted on the
bodies through modification of circulation, downwash effects, pressure
redistribution, and strength of separation.
%

Recent studies of WMLES in transient 3-D channel flows include the
works by \citet{Wiart2018}, \citet{Yang2019} and \citet{Bae2018b}.
\citet{Wiart2018} investigated the performance of WMLES in an ample
set of cases including acceleration in the streamwise direction, and
showed that WMLES is capable of predicting the wall stress with a
reasonable degree of accuracy. \citet{Yang2019} also attained good
results using wall modelling via physics-informed neural networks,
while \citet{Bae2018b} employed a novel parameter-free dynamic wall
model to predict the wall stress in a flow configuration similar to
the present set-up.

\subsection{Wall models}

At the coarse near-wall grid resolutions of WMLES, the usual no-slip
condition ceases to produce an accurate estimate of the momentum drain
at the wall. Hence, wall models are responsible for estimating the
wall-shear stress. The LES equations are integrated in time using the
wall-shear stress provided by the wall model as a Neumann boundary
condition instead of the no-slip condition.  The kinematic
no-penetration condition is maintained for the impermeable walls of
the channel. Three wall models are investigated in the present work:
the equilibrium wall model by \cite{Kawai2012}, and the
non-equilibrium wall models by \cite{Park2014} and \cite{Yang2015}. We
briefly summarise the main characteristics of each model and the
modifications performed in the present work with respect to their
original formulations.

The model by \cite{Yang2015} accounts for non-equilibrium effects
while retaining a moderate complexity. This model assumes a parametric
velocity profile in the near-wall region, where the coefficients are
determined by enforcing a set of physical constraints. These include
the continuity of the profile, the LES matching condition at a
specified wall distance, and the compliance with the vertically
integrated momentum equation, among others.  The model is usually
referred to as integral wall model (IWM), since the momentum integral
constraint is crucial in accounting for non-equilibrium effects.  In
the original formulation, the wall-model input is averaged in time to
regularise the wall-shear stress, which otherwise was found to cause
numerical instabilities.  In the present study, given the
statistically unsteady nature of the flow, the time averaging is
replaced by spatial averaging along wall-parallel planes.  To comply
with the outer LES equations, we modify the original formulation by
\cite{Yang2015} to account for the spanwise pressure forcing.

The non-equilibrium wall model by \cite{Park2014, Park2016} solves the
full Reynolds-averaged Navier-Stokes equations on a separate near-wall
mesh with a mixing-length type eddy-viscosity closure which
dynamically accounts for the resolved portion of the turbulence in the
wall-model domain. This formulation is the most comprehensive amongst
the considered wall-stress models, and accounts for non-equilibrium
effects embedded into the original Navier-Stokes equations. Herein,
this model is termed NEQWM. In order to avoid the skin-friction over
prediction, the resolved turbulent stress is evaluated on the fly, and
it is then subtracted from the modelled stress. Similarly to the IWM,
the formulation by \cite{Park2014} was adjusted to account for the
spanwise pressure forcing.  This turned out to be particularly
important in order to provide the required dominant balance in the
momentum conservation equation for the initial times of the transient.

Lastly, the equilibrium wall model (EQWM) of \cite{Kawai2012} is
derived from the NEQWM by retaining only the wall-normal diffusion
terms.  The model involves a simple ordinary differential equation,
which is solved along the wall-normal direction on each wall face
\citep{Wang2002}. Consistent with the one-dimensional nature of the
model, the spanwise mean pressure gradient vector was projected to the
local flow direction at the matching location, and this was added to
the EQWM equation as a momentum source term. A similar term was added
to the energy equation for consistency.


\subsection{Numerical set-up}

The codes used for wall-modelled calculations are different from the
solver presented for DNS, mainly because the wall models were
conveniently available in other well-validated LES codes.  The
calculations using the NEQWM and EQWM are conducted using the code
CharLES, which is an unstructured-grid finite-volume LES code for
compressible flows developed at the Center for Turbulence Research and
currently maintained by Cascade Technologies, Inc.  The nominal
spatial accuracy of the code is second order, but the reconstruction
scheme upgrades to a fourth-order accuracy on uniform Cartesian grids
\citep{Herrmann2010,Khalighi2011}. The dynamic Smagorinsky model
\citep{Moin1991, Lilly1992} is used as subgrid-scale (SGS) model in
the filtered conservation equations.  The bulk Mach number is fixed at
0.2 for comparison with the incompressible DNS solution.

For the IWM, we use the LESGO solver \citep{LESGO}.  The code solves
the incompressible filtered Navier-Stokes equations in half channel
with a staggered grid, using a pseudo-spectral approach in the
wall-parallel directions and a second-order central finite difference
scheme in the wall-normal direction.  The scale-dependent
Lagrangian-dynamic Smagorinsky model is used as SGS model
\citep{Bou-Zeid2005}.
  
The LES grid resolution is uniform in the three spatial directions and
equal to $(\Delta_1^+, \Delta_2^+, \Delta_3^+)$ = (180, 60, 133) or
$(\Delta_1^*, \Delta_2^*, \Delta_3^*)$ = (0.2, 0.06, 0.14).  The size
of the computational domain is $(L_1^*, L_2^*, L_3^*) = (8\pi, 2, 3\pi
)$, which yields a total of 265,980 grid cells distributed as $(N_1,
N_2, N_3) = (130, 31, 66)$, in the streamwise, wall-normal, and
spanwise directions, respectively.  The internal grids for EQWM and
NEQWM have 30 to 40 cells stretched along the wall-normal
direction. Additionally, the NEQWM shares the same wall-parallel
resolution as the LES grid.  The wall-normal exchange between the wall
model and the LES is located at the centroids of the third grid cell
away from the wall, $x_2^* \approx 0.16$.

The calculations are initialised with a 2-D channel flow in the
statistically steady state at $Re_\tau \approx 1000$.  Then, a
spanwise pressure gradient of $\Pi =10$ is applied to induce a
cross-stream shear layer as in \S\ref{sec:numerics}.  The transverse
mean pressure gradient selected is relatively low in order to mimic
the fact that at high Reynolds numbers, the near-wall layer is in a
quasi-equilibrium state as discussed in \S\ref{subsec:regimes}. The
simulations are run for one eddy-turnover time based on the streamwise
friction velocity, $t^* \approx 1$.

\subsection{Results and discussion}

\begin{figure}

\begin{center}
   \vspace{0.1cm}
 \includegraphics[width=1\textwidth]{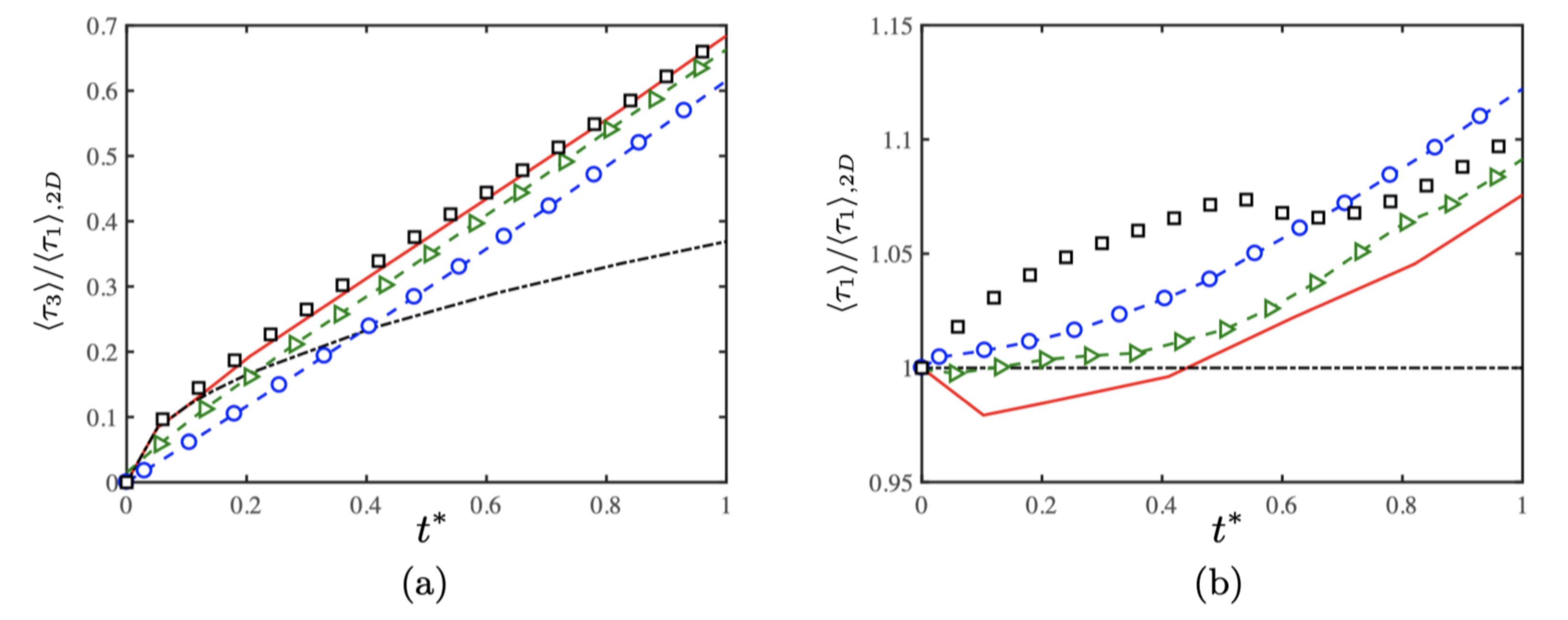}
 \end{center}
\caption{Time evolution of (a) the mean spanwise wall-stress $\langle
  \tau_{3} \rangle$ and (b) the mean streamwise wall-stress $\langle
  \tau_{1} \rangle$ for WMLES and DNS. The stresses are normalised by
  the value of the initial 2-D streamwise wall-stress $\langle
  \tau_1\rangle_{,2D}$.  The lines and symbols are
  (\textcolor{red}{\solid}), DNS; 
  (\squar), WMLES (NEQWM);
  (\trian), WMLES (IWM); 
  (\circle), WMLES (EWQM); 
  (\dchndot), laminar solution. Note that the variations in the vertical axis of
  panel (b) are up to 70\%, while those in panel (b) are only up to
  15\%.
\label{tau1_tau3_wmles}}
\end{figure}

Figure \ref{tau1_tau3_wmles} shows the time evolution of the
streamwise and spanwise mean wall-stress components denoted by
$\langle \tau_1 \rangle$ and $\langle \tau_3 \rangle$, respectively.
We discuss first the predictions for $\langle \tau_3 \rangle$. A
general observation from figure \ref{tau1_tau3_wmles}(a) is that the
NEQWM produces a fairly accurate prediction of $\langle \tau_3
\rangle$ throughout the transient.  For short times ($t^* < 0.1$), the
NEQWM predictions are closely followed by those from IWM, while the
EQWM results in 50\% to 25\% under-prediction of $\langle
\tau_3\rangle$ throughout the initial transient.  The EQWM and the IWM
still capture correctly the growth rate of $\langle \tau_3\rangle$ for
$t^* \gtrsim 0.1$. For $t^* \approx 1$, the errors by NEQWM and IWM
are roughly 2\%, whereas the error for the EQWM is 10\%.  As a
reference, the laminar response of the flow is also included in the
figure, which shows that the spanwise wall stress agrees with the
laminar solution for $t^* < 0.1$.
%

The time evolution of $\langle\tau_1\rangle$ is plotted in figure
\ref{tau1_tau3_wmles}(b).  Note that the variations in
$\langle\tau_1\rangle$ are only up to 10\% and well below the changes
undergone by $\langle\tau_3\rangle$, which are up to 70\%. The EQWM
and the IWM predict the wall-stress throughout the transient within
5\% and 2\% error, respectively. The NEQWM predicts a relatively
faster variation in $\langle \tau_1\rangle$ for $t^* \lesssim 0.4$
compared to IWM and EQWM, with deviations from the DNS up to 7\%. In
all cases, the errors decay as time advances.  As expected, none of
the wall-models is able to reproduce the initial reduction in $\langle
\tau_1 \rangle$ for $t^* \lesssim 0.4$.  Such a decrease in the
streamwise wall-stress component is the result of the complex flow
dynamics discussed in \S\ref{sec:analysis}.  The wall-models
investigated are based on the eddy viscosity assumption; increasing
shear rates in the flow results in additional strain rates. Hence, it
comes as no surprise that WMLES consistently exhibits an approximately
monotonic increase in $\langle \tau_1 \rangle$ after the sudden
spanwise pressure gradient is applied due to the additional transverse
straining of the flow in the near-wall region.

\begin{figure}
\begin{center}
   \vspace{0.1cm}
 \includegraphics[width=1\textwidth]{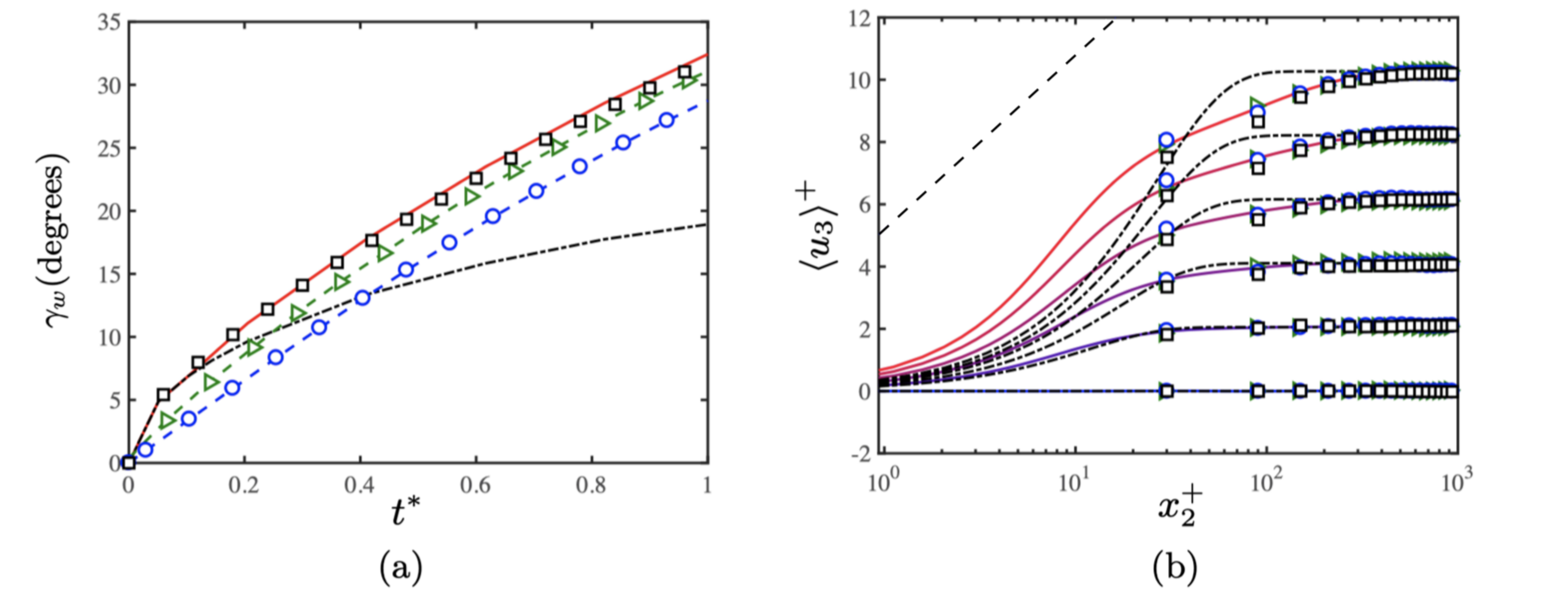}
\end{center}
\caption{(a) Time evolution of the wall shear-stress angle, $\gamma_w
  = \text{tan}^{-1}(\langle \tau_3\rangle/\langle \tau_1\rangle)$. (b)
  Mean spanwise velocity at $t^* = $ 0, 0.21, 0.405, 0.615, 0.825, and
  1.02 (from bottom to top).  The lines and symbols are
  (\textcolor{red}{\solid}), DNS; (\squar), WMLES (NEQWM); (\trian),
  WMLES (IWM); (\circle), WMLES (EWQM); (\dchndot), laminar solution;
  (\dashed), standard logarithmic law, $\langle u_3 \rangle^+ =
  1/0.41\text{ln}(x_2^+) + 5.2$.
\label{alpha_W_wmles}}
\end{figure}

Time evolution of the wall-shear angle, defined as $\gamma_w =
\tan^{-1}(\langle\tau_3 \rangle/ \langle\tau_1\rangle)$, is shown in
figure \ref{alpha_W_wmles}(a).  The performance of the wall models
resembles the trends reported for $\langle \tau_3\rangle$.  This
similarity is easily understood by noting that the relative
time-variations in $\langle \tau_1\rangle$ are modest compared to the
variations in $\langle \tau_3\rangle$.  The development of the mean
spanwise velocity over one eddy-turnover time is shown in figure
\ref{alpha_W_wmles}(b).  All the WMLESs considered provide an
excellent prediction of the boundary layer growth.  The spanwise
velocity profile develops its own logarithmic region for $t^* > 0.6$,
although the slope is substantially smaller than that of equilibrium
channel flows.  The agreement is observed in the turbulent flow
region, where contributions from the SGS models and the wall-models
are expected to play a role in the mean spanwise momentum balance.
These findings highlight the capability of current WMLES and SGS
models to predict the mean spanwise velocity profile that arises in
response to mild transverse pressure perturbations.  Although not
shown, the mean streamwise velocity undergoes only minor changes in
time from its initial 2-D state, and good agreement is also found
between DNS and WMLES.
%

In summary, our results show that the expected errors in WMLES of
non-equilibrium 3-D flows are reduced for increasing degree of
modelling complexity. However, factors such as intricacy in model
implementation or computational cost can favour the adoption of the
simplest wall-models, especially for flow configurations with only
moderate non-equilibrium 3-D effects.

\section{Conclusions}\label{sec:conclusions}

Non-equilibrium turbulent boundary layers with mean-flow
three-dimensionality (3DTBL) are ubiquitous in engineering and
geophysical flows. Under these conditions, the flow is known to
exhibit counter-intuitive behaviours such as Reynolds stress
depletion, wall friction reduction, and misalignment of the Reynolds
stress vector and mean shear, among others. However, previous studies
on 3DTBL have been hampered by their low Reynolds numbers due to
computational constrains, or by the scarcity of time-resolved 3-D data
in experimental studies. In the present work, we have investigated the
transient response of the tangential Reynolds stress in a turbulent
boundary layer with 3-D mean velocity under non-equilibrium
conditions. We have focused our analysis on the multiscale response of
the self-similar momentum-carrying eddies in the flow, which is the
scenario expected at the Reynolds numbers encountered in real-world
applications.

We have performed a series of DNS of fully-developed incompressible
turbulent channel flow subjected to a sudden spanwise mean pressure
gradient. A variety of spanwise to streamwise mean pressure ratios
have been considered ranging from $\Pi=1$ to 100.  The sudden
imposition of the forcing is followed by a continuous change of the
mean-flow magnitude and direction, in which 3-D non-equilibrium
effects prevail. The present set-up is one of the simplest flows
enabling the study of 3-D non-equilibrium wall turbulence, while
maintaining homogeneity in the streamwise and spanwise directions.  We
have considered two moderately high Reynolds numbers, namely
$Re_\tau\approx500$ and $Re_\tau\approx1000$, to uncover the scaling
properties of 3DTBL.

Non-equilibrium effects are observed in both the original frame of
reference as well as in the time- and wall-normal-dependent frame of
reference aligned with the mean shear. The non-equilibrium response of
the flow is controlled by the two non-dimensional parameters of the
problem, namely $Re_\tau$ and $\Pi$.  By assuming that wall
turbulence can be apprehended as a multiscale collection of
wall-attached momentum-carrying eddies with sizes and lifetimes
proportional to $x_2$ and $x_2/u_\tau$, respectively, we have
established that the maximum depletion of the tangential Reynolds
stress is proportional to $\Pi x_2^*$. Therefore, larger eddies are
more prone to experience non-equilibrium effects than the smaller
eddies closer to the wall.  Accordingly, the flow can be classified
into three distinctive flow regimes. For $\Pi < \mathcal{O}(1)$, the
sudden spanwise pressure gradient is too modest to alter the
statistical equilibrium of the momentum carrying eddies.  Conversely,
for $\Pi > 0.03 Re_\tau$ the imposed mean spanwise pressure gradient
is strong enough to leave out of equilibrium eddies at all the scales
across the boundary layer, i.e., from the smallest buffer-layer eddies
up to the very large scale motions populating the outer region. For
$\mathcal{O}(1) < \Pi < 0.03 Re_\tau$, the boundary layer attains an
intermediate state in which eddies closer to the wall evolve in
quasi-equilibrium, whereas eddies further from the wall are influenced
by the non-equilibrium effects.

We have examined the time history of the tangential Reynolds stress
for cases in the fully non-equilibrium regime. The momentum-carrying
eddies undergo an ordered response in time: first, the smallest eddies
(closer to the wall) reduced their Reynolds stress contribution,
followed by the larger eddies (farther from the wall), and so forth.
During the initial transient, the results collapse across several
wall-normal distances and the two Reynolds numbers inspected when the
Reynolds stress drop is assumed to be proportional to $\Pi x_2^*$ and
the time is scaled by $x_2/u_\tau$, consistent with the multiscale
population of eddies discussed above.  The collapse is further
improved for longer times by noting that the characteristic
equilibrium velocity and time scales ($u_\tau$ and $x_2/u_\tau$,
respectively) are no longer representative of eddies in a
non-equilibrium state, which are instead controlled by the
local-in-time scales $u_\tau^\star(t)$ and $x_2/u_\tau^\star(t)$.  Our
results unveil for the first time the self-similar response of
non-equilibrium 3DTBL at high Reynolds numbers and provides the
appropriate scaling framework for future flow comparisons.

We have proposed a structural model for non-equilibrium 3DTBL rooted
on the insight obtained from the physical analysis of the flow. The
model comprises streamwise rolls and streaks at different scales which
are initially in the statistically equilibrium state. The imposition
of the mean spanwise pressure gradient results in the relative
misalignment between the core of rolls and the flow underneath, which
leads to a less efficient configuration of the Reynolds stress
production.  The formulation of the model is consistent with the
self-similar nature of the eddy response, and explains in a
comprehensive manner the findings reported above. The scenario
promoted here is supported by DNS results of the averaged velocity
field conditioned to regions of intense Reynolds stress, which
corroborate the loss of coherence of the layer underneath the core of
the rolls.  Hence, the new structural representation of the flow
entails a quantitative advance of the current leading theories on
transversely strained boundary layers.  Careful inspection of the
Reynolds stress budget reveals that the effect of pressure-strain
correlation is key in the reduction of Reynolds stress within the
additional spanwise shear layer, and that this is the case for all
wall-normal heights.

Finally, the predictive capabilities of three state-of-the-art LES
wall-modelling techniques have been assessed for 3-D channel flows at
$Re_\tau\approx 1000$ and $\Pi=10$.  The models investigated are the
equilibrium wall model by \cite{Kawai2012} (EQWM), and the
non-equilibrium wall models by \cite{Park2014} (NEQWM) and
\cite{Yang2015} (IWM).  As expected, wall models with a higher degree
of complexity yield more accurate predictions of the mean wall-shear,
although the overall performance of the three models is similar.  For
short times, the NEQWM yield the best prediction of the magnitude of
the spanwise wall-shear and the angle of the mean wall stress vector.
The prediction by IWM and EQWM follow in accuracy those by NEQWM. The
larger deviations between wall models are obtained during the early
times of the transient ($t^*<0.1$), while the three models are in
relatively good agreement with the DNS results for longer times
($t^*>1$). Unsurprisingly, none of the wall models considered is able
to account for the initial deficit of Reynolds stress and drag
reduction due to the nature underpinning their eddy-viscosity
formulation.  We have argued that the near-wall layer remains in a
quasi-equilibrium state at high Reynolds numbers, which explains the
fair performance of WMLES based on equilibrium assumptions in
transient 3-D boundary layers.

\section*{Acknowledgements} 

This investigation was funded by the Office of Naval Research (ONR),
Grant N00014-16-S-BA10. The authors acknowledge the assistance of
Prof. Xiang I.A. Yang in performing the wall-modelled simulations with
IWM.




\bibliographystyle{jfm} 
\bibliography{3DTBL}

\end{document}

%% file: macros.tex
\usepackage{graphicx,color}
\usepackage{amssymb,latexsym}
\usepackage{url}

\def\xvec{\boldsymbol{x}}

\def\spacce#1{\hskip #1pt}
\def\drawline#1#2{\raise 2.5pt\vbox{\hrule width #1pt height #2pt}}
\def\solid{\drawline{24}{.5}\nobreak}

\def\bdash{\hbox{\drawline{5.8}{.5}\spacce{2}}}

\def\dashed{\bdash\bdash\bdash\nobreak}

\def\bdot{\hbox{\drawline{1}{.5}\spacce{2}}}

\def\dotted{\hbox{\leaders\bdot\hskip 24pt}\nobreak}

\def\dchndot{\hbox%
{\drawline{4.6}{.5}\spacce{2}\drawline{1}{.5}\spacce{4.6}\drawline{4.6}{.5}\spacce{2}\drawline{1}{.5}}\nobreak }

\def\circle{$\circ$\nobreak }

\def\trian{\raise 1.25pt\hbox{$\scriptstyle\triangle$}\nobreak}

\def\dtrian{\raise 1.25pt\hbox%
{$\scriptscriptstyle\bigtriangledown$}\nobreak}

\def\squar{\raise 1.25pt\hbox{$\scriptstyle\Box$}\nobreak}

\def\diamon{\raise 1.25pt\hbox{$\scriptstyle\diamond$}\nobreak}

\def\beq{\begin{equation}}
\def\eeq{\end{equation}}
\def\la{\label}

\def\ggg{{\it g}}

